\documentclass[fleqn,usenatbib]{mnras}

\usepackage{newtxtext,newtxmath}

\usepackage{color}
\usepackage[T1]{fontenc}

\usepackage{subfigure}
\usepackage{enumerate}

\DeclareRobustCommand{\VAN}[3]{#2}
\let\VANthebibliography\thebibliography
\def\thebibliography{\DeclareRobustCommand{\VAN}[3]{##3}\VANthebibliography}

\usepackage{graphicx}	% Including figure files
\usepackage{amsmath}	% Advanced maths commands
\title[Scatter in the $M_{\star}$-$M_{\mathrm{BH}}$ relation]{Different physical and numerical sources of scatter in the $M_{\mathrm{BH}}$-$M_{\star}$ relation and their connection to galaxy evolution}

\author[B.~Zhu and V.~Springel]{Bocheng Zhu$^{1,2}$ and Volker Springel$^{2}$\thanks{\href{mailto:vspringel@mpa-garching.mpg.de}{vspringel@mpa-garching.mpg.de}}
\vspace*{0.1cm}\\%
$^{1}$Key Laboratory for Computational Astrophysics, National Astronomical Observatories, Chinese Academy of Sciences, Beijing, China\\%
$^{2}$Max-Planck-Institut für Astrophysik, Karl-Schwarzschild-Straße 1, 85741 Garching, Germany}

\date{Accepted XXX. Received YYY; in original form ZZZ}

\pubyear{2024}

\begin{document}
\label{firstpage}
\pagerange{\pageref{firstpage}--\pageref{lastpage}}
\maketitle

\begin{abstract}
Observations have established that the masses of supermassive black holes (SMBHs) correlate tightly with the stellar masses of their host galaxies, albeit with substantial scatter. The {magnitude} of this scatter as a function of galaxy mass and redshift contains valuable information about the origin of SMBHs and the physical nature of their co-evolution with galaxies. In this work, we highlight this connection by studying the scatter in the $M_{\mathrm{BH}} - M_{\star}$ relation for massive galaxies in the Illustris, {TNG100}, and EAGLE cosmological simulations. We find that TNG100 shows significantly lower scatter than Illustris and EAGLE, reflecting different BH feedback models. Using numerical experiments, we separate different contributions to the scatter, including an intrinsic component. At $z=0$, Illustris and EAGLE show $\sim0.3$ dex intrinsic scatter dominated by BH accretion, while the smaller scatter in TNG100 is dominated by hierarchical merging, implying more tightly quenched massive galaxies. BH seed mass variations can add scatter, though their impact at $z=0$ depends on the feedback model. Without AGN feedback the scatter is much larger for low-mass galaxies ($\gtrsim0.5$ dex for $\log M_{\star}<10^{10.5},{\rm M_{\odot}}$ at $z=0$–3), underscoring the crucial role of feedback in SMBH–galaxy co-evolution. In contrast, hierarchical merging of quenched systems is the main factor reducing scatter for massive galaxies. Based on our results, we expect that the scatter in the $M_{\mathrm{BH}} - M_{\star}$ relation at high redshift could be particularly powerful in providing clues to the origin of SMBHs.
\end{abstract}


%%%%%%%%%%%%%%%%%%%%%%%%%%%%%%%%%%%%%%%%%%%%%%%%%%

%%%%%%%%%%%%%%%%% BODY OF PAPER %%%%%%%%%%%%%%%%%%

\section{Introduction}

It is widely accepted that every massive galaxy hosts a supermassive black hole (SMBH) at its center \citep{magorrian98, genzel2010}. Observations further find that there exist tight correlations between the properties of SMBHs and their host galaxies \citep{magorrian98, FM2000, Gebhardt2000, tremaine2002, haring04,kormendy2013,reines2015,terrazas2016}, such as the correlation between stellar mass and central black hole (BH) mass of galaxies, $M_{\mathrm{BH}}- M_{\star}$. These tight correlations are expected to contain important information about galaxy evolution. However, how these correlations form, and whether they still exist at high redshift remains under debate.

There are two main explanations for the $M_{\mathrm{BH}} - M_{\star}$ relation. Some works suggest that the relation may simply arise as a regression to the mean when many mergers occur \citep{peng2007, hirschmann2010, Jahnke2011, angles-alcazar2013}. \citet{peng2007} use a simple random-merging model to study the formation of the relation between the black hole (BH) mass $M_{\rm BH}$ and stellar mass $M_{\star}$. They find that the relation between $M_{\rm BH}$ and $M_{\star}$ will become tight, with a logarithmic slope of close to unity after experiencing several mergers, regardless of the initial distribution of the BH and stellar mass. \citet{hirschmann2010} extend Peng's work to different models and find that the conclusion still holds in the $\Lambda$CDM cosmogony. \citet{Jahnke2011} further study the origins of the scaling relation between $M_{\rm BH}$ and $M_{\star}$ within the $\Lambda$CDM model including also BH accretion and star formation, confirming that the scaling relation can be established by hierarchical merging, despite the presence of star formation and BH accretion.

Previous analyses of observations have also shown that the establishment of the $M_{\rm BH}$ and $M_{\star}$ relation cannot be accounted for by mergers alone \citep{lahav2011,ginat2016}. An arguably more popular explanation for the tight correlation between $M_{\rm BH}$ and $M_{\star}$ is that there is a co-evolution of SMBHs and their host galaxies \citep{kormendy2013, heckman2014, sun2015, zhuang2023} in which they mutually influence each other in some way. Over the past decades, many theoretical works focused on self-regulated growth mechanisms of the stellar content of galaxies and their SMBHs in which feedback processes play a critical role \citep{dmatteo05, hopkins06,  booth09, debuhr11, weinberger17}. Early works like \citet{dmatteo05} proposed a physical picture where the  $M_{\rm BH} -M_{\star}$ relation is a direct consequence of the self-regulation of BH accretion by  AGN feedback, triggered by galaxy mergers. Other works like \citet{Silk1998} and \citet{zinger20} propose a related scenario where the causal origin of the tight relation between SMBHs and their host galaxies originates {from} the competition between the cumulative energy released from SMBHs and the potential binding energy of their host halo.

Advances in cosmological hydrodynamical simulations have allowed studies of the relation between SMBHs and their host galaxies in the cosmological context, including a modeling of astrophysical processes like BH accretion and star formation. In current cosmological simulations, the relation between BHs and their galaxies has in fact become a baseline for testing the validity of the underlying galaxy formation model \citep[][etc.]{vogelsberger14, schaye15, weinberger17, dave19, parkmor2023}. \citet{habouzit2021} have studied the $M_{\rm BH}-M_{\star}$ relation in different recent cosmological simulations and found that all of them can roughly reproduce the relation observed for massive galaxies. Also, \citet{liyuan2020} analysed the properties of BHs and their host galaxies in Illustris and IllustrisTNG, finding that the $M_{\rm BH} -M_{\star}$ relation has a slight offset in these two simulations compared to observations, {and is also tighter than observed.}
%and is tighter as them as well.

Previous studies of the relation between BHs and their host galaxies mainly focused on the relation itself, whereas few works considered the scatter of the relation as well. The findings of \citet{liyuan2020} however suggest that the scatter of the relation may contain fruitful information about galaxy evolution. If the relation can either arise through hierarchical merging or through co-evolution, then the scatter of the relation should mainly come from these two processes. For the former process the scatter reflects the degree to which galaxies do not yet strictly lie on the relation as a result of a limited number of mergers, while the latter process can introduce -- or even enlarge -- the scatter when the star formation rate and BH accretion rate are not strictly proportional to each other despite the presence of feedback. Aside from these astrophysical processes, we note that a potential variation of the BH seed masses at high redshift, which is related to the origin of SMBHs, can also introduce additional scatter in the relation.

In cosmological simulations, the scatter in the $M_{\rm BH}-M_{\star}$ relation may not purely come from physical origins. Because the gravitational and hydrodynamical processes governing the evolution of galaxies are highly non-linear, simulations exhibit aspects of dynamical chaos \citep{keller19, genel19, borrow23b}. Extremely small variations in any resolution element can quickly grow and turn into sizable differences. This can happen even for identical initial conditions, because variations in numerical round-off from mathematically allowed re-orderings of the operations in parallel codes are sufficient to introduce such small differences.  In addition, {many current cosmological simulations resort to stochastic models to address certain subgrid physics. For example, decisions such as whether a star particle should form, or whether stellar or BH feedback is triggered, are made based on drawing random numbers.} The exact sequence of random numbers used then becomes a source of numerical variations, and can thus contribute scatter to the $M_{\rm BH}-M_{\star}$ relation that is of purely numerical origin.

Several works have demonstrated these numerical variations explicitly.  For example, \citet{keller19} performed a set of cosmological zoom-in simulations with different galaxy formation models but identical initial conditions, and found that the scatter in the stellar masses due to numerical variations and stochastic modeling grows with a Lyapunov rate, but then saturates due to stellar feedback. \citet{genel19} examined properties of galaxies in the cosmological simulations carried out with the physics model and code {of IllustrisTNG}, and also found similar trends. \citet{borrow23b} investigated variations due to numerics and stochastic modeling in cosmological simulations with the EAGLE model and confirmed that the random variability should be taken into account when interpreting results for individual galaxies.

In this work, we investigate the scatter of the $M_{\rm BH}-M_{\star}$ relation and its possible connection to galaxy evolution physics in the Illustris, IllustrisTNG, and EAGLE simulations.  To improve the statistics of the IllustrisTNG results, we also consider the large volume MillenniumTNG simulation box, whose physics model is otherwise however identical to TNG. We try to quantify the different origins of the scatter in these cosmological simulations with the help of several numerical experiments, and compare the implied intrinsic scatter predicted by the cosmological simulations with observations.

The paper is organized as follow. In Section~\ref{sec:scatter_in_diff_cosmol_sim}, we briefly introduce the Illustris, IllustrisTNG, EAGLE, and MillenniumTNG simulation projects. The scatter of the $M_{\rm BH}-M_{\star}$ relation for massive galaxies seen in these simulations, and the impact of galaxy selection criteria and numerical resolution is also briefly discussed. In Section~\ref{sec:origin}, we introduce possible origins for the scatter and analyze the data from numerical toy experiments to constrain the numerical contributions to the scatter in the model {of IllustrisTNG}. We also apply the derived formalism to other cosmological simulations and estimate their intrinsic scatter and compare it with observations. In Section~\ref{sec:dicussion}, we discuss further aspects of galaxy evolution from the viewpoint of the scatter in the $M_{\rm BH}-M_{\star}$ relation, in particular how this relates to galaxy quenching and to overmassive BHs at high redshift. Finally, we summarize and discuss our results in Section~\ref{sec:conclusion}.

\section{Scaling relation and scatter in different cosmological simulations}\label{sec:scatter_in_diff_cosmol_sim}

In this section we introduce the different cosmological simulations we analyze and we present basic measurements of the scatter in their predicted $M_{\rm BH}-M_{\star}$ relations. Note that although the relation between BH mass and central stellar velocity dispersion ($M_{\mathrm{BH}}-\sigma$) is somewhat tighter observationally, we here focus on the $M_{\rm BH}-M_{\star}$ relation for practical reasons, as it can be straightforwardly and without ambiguity measured from the simulations, whereas the central velocity dispersion of stars can be sensitive to the spatial region over which it is averaged, and it is also not directly given in the public data releases of all the cosmological simulations.

\subsection{Simulations}

In this work, we have selected the Illustris, IllustrisTNG, and EAGLE simulations for comparing their $M_{\rm BH}-M_{\star}$ relations, and in particular the scatter they exhibit around this relation.  We shall also include a few selected measurements from MillenniumTNG, which has the same physics model as IllustrisTNG but larger volume, to test the statistical robustness of our findings for IllustrisTNG. Below we briefly introduce these cosmological simulations, while for a detailed description we refer to the corresponding methods papers.

\subsubsection{Illustris}

The Illustris project is a cosmological simulation with box size equal to $106.5\, {\rm cMpc}$ that was carried out with the TreePM magneto-hydrodynamical moving-mesh code {\small AREPO} \citep{Springel2010}. The mass resolutions of dark matter (DM) particles and gaseous cells are $6.3\times 10^6\,{\rm M_{\odot}}$ and $1.3\times 10^6\,{\rm M_{\odot}}$, respectively{, where the gas resolution corresponds to the target cell mass}. The subgrid models for BH seeding, accretion and feedback, and the star formation prescription and stellar feedback applied in Illustris are described in \citet{vogelsberger14}. We concisely summarize these subgrid treatments here. For more details, please refer to the original paper.

Black holes are modelled as sink particles at the center of galaxies. BH seeds with mass $m_{\rm seed}=1.42\times 10^5\, {\rm M_{\odot}}$ are placed in all FOF halos with mass higher than $7.1\times10^{10}~{\rm M_{\odot}}$ and which do not contain a BH yet{, by converting the densest gas cell in the halo into a BH particle}. The BH accretion rate is estimated with the Bondi–Hoyle–Lyttleton formula subject to an Eddington limit: \begin{equation}
    \begin{aligned}
     \dot{M}_{\rm BH} = {\rm min}(\dot{M}_{\rm Edd}, \dot{M}_{\rm Bondi})\hspace*{3.3cm}\,\\
     ={\rm min}(\alpha 4 \pi G^2 M^2_{\rm BH}\bar{\rho}/\bar{c_{\rm s}}^3,4\pi G M_{\rm BH} m_{\rm p}/\epsilon_{\rm r} \sigma_{\rm T} c) ,
     \end{aligned}
\end{equation}
where a boost factor $\alpha=100$ was adopted, which was first introduced in \citet{springel05BHFBModel} as a correction for the unresolved multiphase nature of the gas surrounding the BH. $\bar{\rho}$ and $\bar{c_{\rm s}}$ are the kernel-weighted density and sound speed around the BH, respectively.

The AGN feedback incorporated in Illustris distinguishes two modes, depending on how the accretion rate compares to a transition threshold of $\dot{M}_{\rm BH, trans}=0.05\dot{M}_{\rm Edd}$. At higher accretion rate than this threshold, the feedback model is based on the `quasar-mode' model of \citet{springel05BHFBModel}. The feedback energy is deposited around the BH in the form of thermal energy, with a total energy release of $0.05\dot{M}_{\rm BH} c^2$. At low accretion rate, the feedback model is based on the `radio-mode' description of \citet{sijacki07}. Here the feedback energy is released in a randomly selected region within $\sim100\, {\rm kpc}$ around the BH to mimic the hot bubbles created by AGN jets. The coupling efficiency for this model is assumed to be 0.35.

The star formation and stellar feedback model incorporated in Illustris is based on a simple two-phase ISM and non-local stellar wind subgrid model first described in \citet{springel03}. The gas in the simulation becomes star-forming when its density is higher than $\sim 0.1\, {\rm cm^{-3}}$, and the temperature is equal or less than $10^{4}\,{\rm K}$. The star-forming gas is assumed to be a two-phase ISM regulated by SN feedback. The thermal state of the star-forming gas is calculated by an effective equation of state based on the two-phase ISM model and a star formation rate parameterization where the gas consumption timescale is assumed to scale with the local free-fall time. For stellar feedback in the form of galactic winds, a non-local momentum injection is adopted which is realized in terms of wind particles that are decoupled from the gas until they leave the star-forming region. The initial stellar wind velocity is prescribed and parameterized as a multiple of the local 1D dark matter velocity dispersion, which serves as a proxy for the escape velocity of the halos. For the detailed parameters of these subgrid models, please refer to \citet{vogelsberger14}.

\subsubsection{IllustrisTNG}

The IllustrisTNG project \citep[hereafter TNG; ][]{marinacci18, naiman18, pillepich18b, springel18, nelson2018}  is a suite of cosmological simulations of galaxy formation also performed with {\small AREPO}. The TNG project employs three different box sizes: $51.7\,{\rm cMpc}$, $110.7\,{\rm cMpc}$, and $302.6\,{\rm cMpc}$ on a side, and the corresponding simulations are referred to as TNG50, TNG100, and TNG300, respectively.  The TNG50 box was carried out at four different resolutions, while the TNG100 and TNG300 simulations have been computed for three different resolutions each. The mass resolutions of the DM particles in the different resolutions of TNG100 are $7.5\times 10^6\,{\rm M_{\odot}}$, $6\times 10^7\,{\rm M_{\odot}}$, and $4.8\times 10^8\,{\rm M_{\odot}}$, while the mass resolutions of the gases cells in the different resolutions of TNG100 are $1.4\times 10^6\,{\rm M_{\odot}}$, $1.1\times 10^7\,{\rm M_{\odot}}$, and  $8.9\times 10^7\,{\rm M_{\odot}}$. Throughout this work, we mainly utilize the data of the TNG100 simulation with the highest resolution, allowing us to best compare with the other simulations, which have similar mass resolution. However, we will also utilize the two TNG100 realizations with lower resolution to investigate the robustness and resolution dependence of our conclusions.

The subgrid models for BHs and stars in TNG100 are quite similar to the ones in Illustris, but differ in a small number of crucial aspects.  We briefly describe these differences here.  For a more detailed description please refer to \citet{weinberger17} and \citet{pillepich18}.

The BH seeding in TNG is similar to Illustris, but the BH seed mass is substantially higher, $1.18\times 10^6\, {\rm M_{\odot}}$, which is around one order of magnitude more massive than in Illustris. The BH accretion rate is still calculated with the Bondi-Hoyle-Lyttleton formula, but now the boost factor $\alpha$ is omitted. The increase in the seed mass compensates for the missing boost in the smallest black holes, so that they still have a chance to grow substantially on a reasonably short timescale. For the AGN feedback, TNG also incorporates a two-mode model. However, the transition accretion rate between two modes scales as a function of the BH mass itself:
\begin{equation}
\dot{M}_{\rm BH}/\dot{M}_{\rm Edd}={\rm min}\left[2\times10^{-3}\left(\frac{M_{\rm BH}}{10^8{\rm M_{\odot}}}\right)^2,0.1\right].
\end{equation}
In addition, the AGN feedback model at low accretion rate (radio-mode) is replaced in TNG by a kinetic feedback prescription where the feedback energy is injected in the form of momentum kicks within a small region around the BH. For a detailed description and motivation of this AGN feedback in TNG, please refer to \citet{weinberger17}.

Finally, for star formation and stellar feedback TNG adopts a broadly similar approach to the one in Illustris. TNG however introduces a lower limit and a redshift-dependence in its parameterization of galactic wind feedback, and a metallicity-dependence of the mass loading factor for the winds, which combined make the stellar feedback more efficient in low-mass galaxies. For a detailed description of the star formation and stellar feedback, and their parameters, please refer to \citet{pillepich18}.

\subsubsection{EAGLE}

The EAGLE project \citep{schaye15,crain15} is a cosmological {simulation} with a box size of $100\, {\rm cMpc}$ performed with the TreePM smoothed particle hydrodynamics (SPH) formulation {\small ANARCHY}, implemented in the {\small GADGET-3} code. The mass resolutions of the DM particles and gas particles are $9.7\times 10^6\,{\rm M_{\odot}}$ and $1.8\times 10^6\,{\rm M_{\odot}}$, respectively. For a full description of the subgrid models for BHs and stars in the EAGLE simulation, please refer to \citet{schaye15}. We here only offer a brief summary.

The BH seeding algorithm in EAGLE is similar to Illustris and TNG. BH seeds with a mass $M_{\rm seed}=1.48\times 10^5\,{\rm M_{\odot}}$ are introduced in DM halos with mass $M_{\rm DM}>1.48\times 10^{10}\,{\rm M_{\odot}}$. The BH accretion rate is calculated with a modified Bondi-Hoyle-Lyttleton formula  following \citet{rg2015}, after taking the angular momentum of the accreting gas into account:
\begin{equation}
\dot{M}_{\rm BH} = {\rm min}\left[\dot{M}_{\rm Edd}, \dot{M}_{\rm Bondi}\times{\rm min}\left((\bar{c_{\rm s}}/V_{\Phi})^3/C_{\rm visc},1\right)\right],
\end{equation}
where $\bar{c_{\rm s}}$ is the kernel-weighted sound speed, $V_{\Phi}$ is the circular speed of the gas around the BH, and $C_{\rm visc}$ is a free parameter representing the viscosity of the accreting gas. The AGN feedback is purely thermal and based on \citet{booth09}. In this model, the AGN feedback energy is injected around the BH with a feedback efficiency of 0.015, but the feedback energy is accumulated before it is injected until the deposited feedback energy {can heat a single gas particle by at least $\Delta T \sim 10^{8.5}\,{\rm K}$.}
%can increase the temperature of the surrounding gas by $\Delta T \gtrsim  10^{8.5}{\rm K}$. 
This is done to avoid that the thermal energy injection can be rendered inefficient by numerical overcooling problems that can arise when the low density of the heated gas can (at least initially) not be spatially resolved.

The star formation model in EAGLE is based on \citet{schaye08}. Star formation will take place when the hydrogen number density of the gas particles is higher than $0.1(Z/0.002)^{-0.64}\,{\rm cm^{-3}}$ {and the gas temperature $T$ satisfies $\log T<\log T_{\rm EOS} +0.5$. The temperature floor, $T_{\rm EOS}$, follows a polytropic relation $P_{\rm EOS} \propto \rho^{4/3}$, normalized to $T_{\rm EOS}=8\times10^3 \,{\rm K}$ at $n_{\rm H}=0.1\, {\rm cm^{-3}}$. } The star formation rate is then calculated by a pressure-dependent formula derived from a Kennicutt–Schmidt law. 
%A power-law equation of state, $p\propto\rho^{4/3}$, is prescribed {, with a temperature floor to prevent artificial fragmentation at the resolution limit,} when the gas particles become star-forming. 
{Feedback from core-collapse supernovae is implemented using a stochastic thermal injection scheme based on \citet{dv2012}. At an age of $3\times10^7$ yr, each newly formed star particle probabilistically heats a subset of its neighbouring gas particles by a fixed temperature increase of $\Delta T \gtrsim 10^{7.5}\,{\rm K}$. This approach ensures that the energy injection is locally effective, avoiding artificial radiative losses due to numerical overcooling.}
%Feedback from supernova is injected with a stochastic model in the form of thermal energy based on \citet{dv2012}. Similar to the treatment of AGN feedback and in order to prevent numerical overcooling problems, the feedback energy is accumulated and only deposited once the energy release can cause a temperature increase of $\Delta T\gtrsim 10^{7.5}{\rm K}$ in the surrounding gas.

\subsubsection{MillenniumTNG}

The recent MillenniumTNG project \citep[hereafter MTNG; see introductory and model descriptions in][]{parkmor2023, mtng6} is an extension of IllustrisTNG with a box size of $740\, {\rm cMpc}$ in its flagship hydrodynamical model (the project also contains very large dark matter simulations not considered here). Several papers have presented first scientific results based on MTNG, among them high-redshift galaxy populations \citep{mtng8}, large-scale galaxy clustering \citep{mtng7}, weak lensing \citep{mtng9}, and intrinsic alignments of galaxies \citep{mtng1}. The MTNG volume is around {300 times larger than that of TNG100, and} 15 times larger than that of TNG300, allowing a much better representation of rarer massive systems. The mass resolution of DM particles and gas particles in MTNG is $1.7\times 10^8\,{\rm M_{\odot}}$ and $3.1\times 10^7\,{\rm M_{\odot}}$, respectively, which is comparable to TNG300, but worse than TNG100.

The subgrid model of MTNG is almost identical to TNG100, only the magnetic fields and the individual metal species have been omitted in MTNG for practical reasons; the associated memory savings were necessary to fit the calculation into the available supercomputer partition. The lack of individual metal species has a negligible influence on the simulation results but the magnetic field does affect the galaxy evolution in subtle ways. The results in \citet{pillepich18} and \citet{parkmor2023} indicate that the magnetic field can enhance the BH accretion since the magnetic pressure partly contributes to the total pressure. Since the total pressure is roughly constant in order to balance the gravitational weight, the necessary thermal pressure is slightly reduced in the TNG simulations compared to MTNG. This makes the accretion rate calculated by the Bondi formula slightly larger in turn, allowing the BHs to grow a bit faster.

\begin{figure*}
   \includegraphics[width=\textwidth]{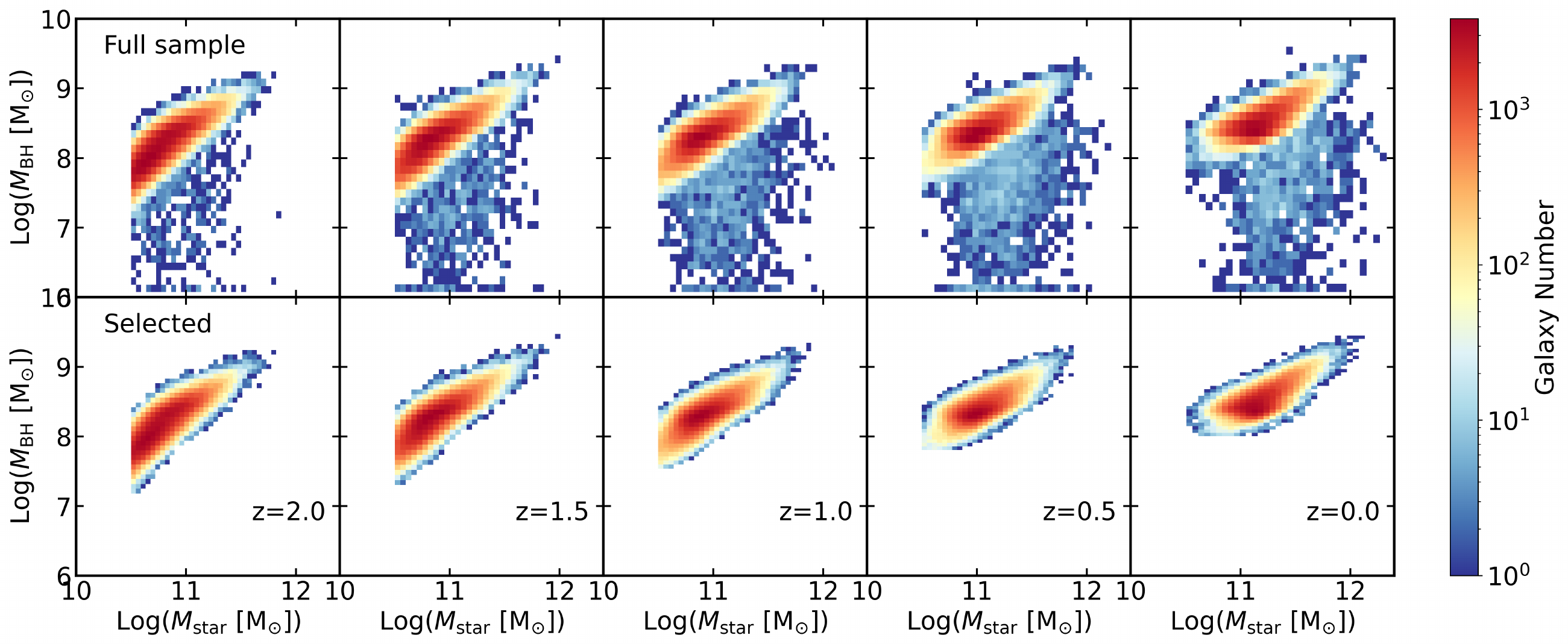}
   \caption{\textit{Upper panels:} The data samples in MTNG from $z=2$ to $z=0$ based on the selection criteria $\sigma>150\, {\rm km\, s^{-1}}$ and   $M_{\star}>10^{10.5}\,{\rm M_{\odot}}$. \textit{Lower panels:} The ``main sequence'' samples in MTNG after discarding outliers with a GMM clustering algorithm. {The figure shows that GMM clustering can successfully remove the subpopulation that does not belong to the main relation. We will use the selected part shown in the lower panels for the analysis in the rest of this work.}}
    \label{mtng_selection}
\end{figure*}

\begin{figure*}
   \includegraphics[width=0.8\textwidth]{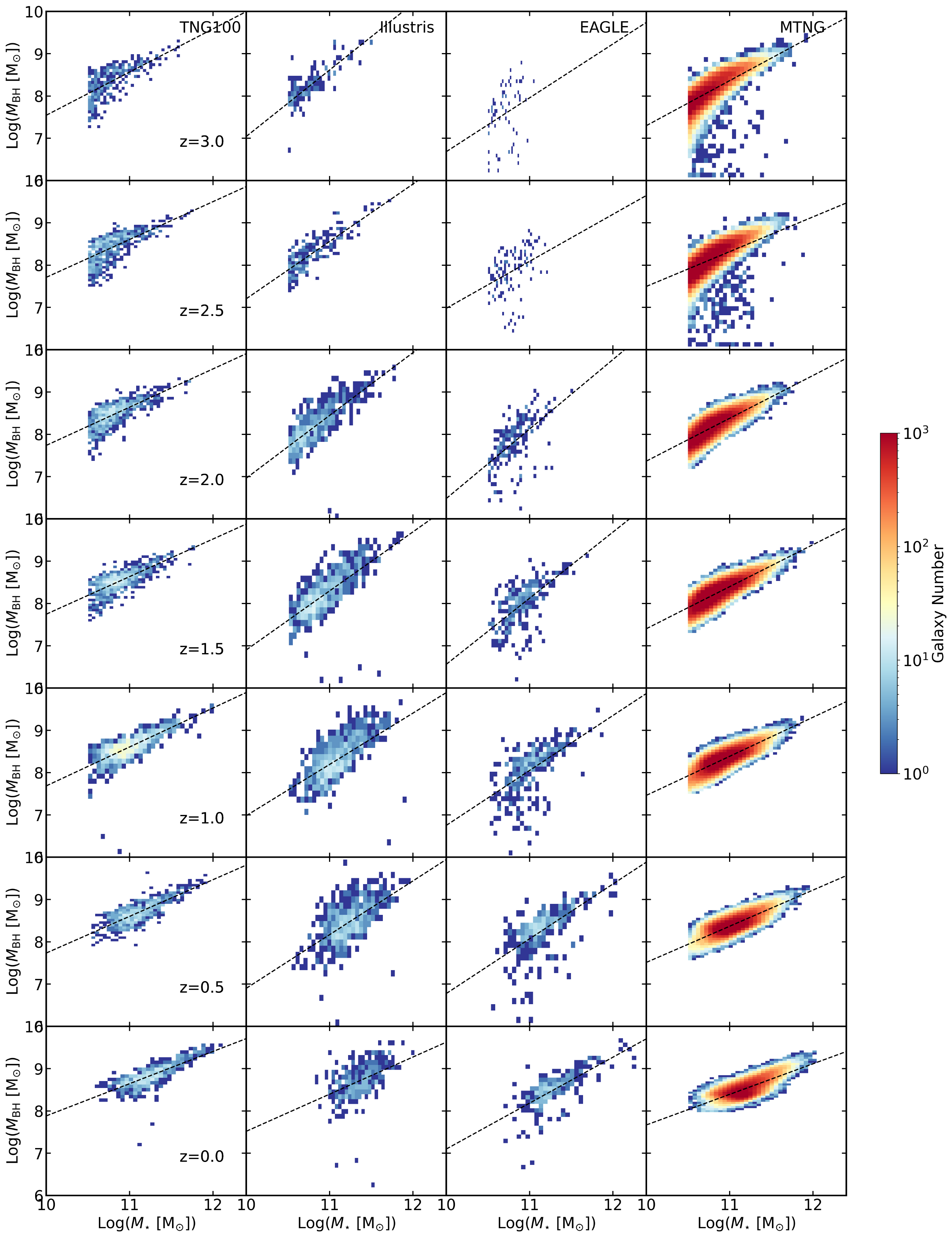}
   %\subfigure{\includegraphics[width=0.45\textwidth]{Figures/CS_select_11.pdf}}
   \caption{The selected data samples in Illustris, TNG, EAGLE and MTNG from $z=3$ to $z=0$, as labeled, with the selection criteria  $\sigma>150\, {\rm km\, s^{-1}}$ and $M_{\star}>10^{10.5}\,{\rm M_{\odot}}$. The dashed lines show the results of our power-law fits to the relations. The aim of this figure is to illustrate the distribution of selected data samples from the Illustris, TNG, EAGLE, and MTNG simulations, spanning redshifts from $z=3$ to $z=0$, with specific selection criteria ($\sigma > 150\, {\rm km\, s^{-1}}$ and $M_{\star} > 10^{10.5}\,{\rm M_{\odot}}$), and the corresponding fitting results.}
    \label{selecteddata}
\end{figure*}

\subsection{Data processing}

In this subsection, we introduce our simulation data processing procedure, in particular with respect to the sample selection and our fitting procedure for the $M_{\rm BH}- M_{\rm \star}$ relation.

\subsubsection{Data selection}\label{sec:selectcriteria}

Since our work focuses on the scatter in the $M_{\rm BH}- M_{\rm \star}$ relation of reasonably massive galaxies, we only select galaxies in the cosmological simulations which have a stellar mass $M_{\star}>10^{10.5}\,{\rm M_{\odot}}$, and which sit in halos with a total velocity dispersion of $\sigma>150\, {\rm km\, s^{-1}}$. To examine the robustness of our results to these selection criteria, we also replaced the stellar mass criterion with $M_{\star}>10^{11}\,{\rm M_{\odot}}$, for comparison.

Throughout this work, we define the BH mass as the sum of the subgrid masses of all BH particles within a given subhalo. In practice, this quantity is typically dominated by a single central BH, so the difference between summing all BH particles and considering only the most massive one is negligible.

For the sample taken from MTNG based on the selection criteria above we find that there is a comparatively prominent subpopulation with lower BH mass compared to the ``main sequence'' of the $M_{\rm BH}- M_{\rm \star}$ relation, as shown in the upper panels of Figure~\ref{mtng_selection}. Most of these galaxies belong to massive galaxy clusters and are not central galaxies, and further investigation reveals that they have a very similar color compared to the galaxies with similar stellar mass on the main sequence. While a deep investigation of their formation history is beyond the scope of the current work, we believe that the much smaller abundance of these systems in TNG100 is related to a known deficiency of its black hole reposition algorithm \citep{Borrow2023}, which can make satellite galaxies in clusters lose their black hole prematurely once all of their gas is stripped, a problem that has been fixed in MTNG.

For simplicity, we here use a Gaussian mixture model (GMM) clustering algorithm to separate these satellite galaxies from the samples taken from MTNG. The bottom panel of Figure~\ref{mtng_selection} shows that the GMM performs well for this task, and so we use the resulting cleaned MTNG-sample for our analyses below. {This selection is motivated by observational limitations: previous observational samples, such as those compiled in \citet{kormendy2013}, were strongly biased against the detection of undermassive black holes in massive host galaxies, particularly satellite systems. For consistency with these observational constraints, we restrict our comparison sample in MTNG to galaxies near the observed $M_{\rm BH}-M_\star$ main sequence.}

\subsubsection{Fitting procedure}\label{sec:fit}

The data samples and the fitting results from $z=3$ to $z=0$ in TNG, Illustris, EAGLE and MTNG with the selection criteria  $\sigma>150\, {\rm km\, s^{-1}}$ and $M_{\star}>10^{10.5}\,{\rm M_{\odot}}$ are presented in Figure~\ref{selecteddata}. For the fitting procedure, we adopt {a linear relationship between the logarithms of the masses} as in \citet{kormendy2013}, in the form
\begin{equation}
\log\left(\frac{M_{\rm BH}}{10^9\,{\rm M_{\odot}}}\right)=k\log\left(\frac{M_{\star}}{10^{11}\,{\rm M_{\odot}}}\right)+b .
\end{equation}
Similar to observational data, which {have} individual measurement errors, the simulation data has also errors that can be identified with the uncertainties originating in numerical variations or stochastic modeling, i.e.~$M_{\rm BH}+\delta_{\rm BH}$ and $M_{\star}+\delta_{\star}$, where $\delta_{\rm BH}$ and $\delta_{\star}$ are variations introduced by the numerical discretization scheme or the use of stochastic subgrid modeling. In this context, these variations are analyzed in logarithmic space, i.e., we consider variations in $\log M_{\rm BH}$ and $\log M_\star$. In the rest of this paper, all scatter/variance values refer to the logarithmic scatter, defined as the standard deviation in $\log M_{\rm BH}$ at fixed $\log M_\star$. Due to the existence of these errors, one should in principle not use  the simulation data with the simple linear regression method as this can yield a biased estimate of the slope. Indeed, if the simple linear regression is used, the estimated slope $\hat{\beta}$ becomes 
\begin{equation}
    \hat{\beta} = \beta \sigma^2/(\sigma^2+\sigma_{\star}^2)=\lambda \beta ,
\label{slopeestimation}
\end{equation}
where $\beta$ is the real unbiased slope, $\sigma$ is the intrinsic scatter of the stellar mass itself, and $\sigma_{\star}$ is the standard deviation of $\delta_{\star}$. Then the scatter $\hat{\epsilon}$ of the $M_{\rm BH}-M_{\star}$ relation obtained with the biased estimation of slope  will become
\begin{equation}
\hat{\epsilon}^2 = \epsilon^2+(1-\lambda)^2\beta^2\sigma^2+\lambda^2\beta^2\sigma_{\star}^2+\sigma_{\rm BH}^2 ,
\label{scatterformula}
\end{equation}
where $\epsilon$ is the unbiased intrinsic scatter of the $M_{\rm BH}-M_{\star}$ relation, and $\sigma_{\rm BH}$ is the standard deviation of $\delta_{\rm BH}$. Equation (\ref{scatterformula}) shows that the biased estimation of the slope will ultimately yield an extra contribution to the estimated scatter of $(1-\lambda)^2\beta^2\sigma^2 + \lambda^2\beta^2\sigma_{\star}^2$, beyond an additive contribution from the stochastic scatter in the black hole masses, $\sigma_{\rm BH}^2$, themselves. Therefore, it is necessary to take special care of the estimation of the slope in case the stochastic numerical errors in the stellar masses are not completely negligible.

Since we do not have any reliable information about the size of the errors due to the numerical variations and stochastic modelling, we utilize a grouping method to obtain an unbiased estimate of slope. The procedure of the grouping method is as follows:
\begin{itemize}
    \item Subdivide the data into different bins based on the stellar mass.
    \item Average the BH mass and stellar mass within each bin.
    \item Apply a weighted least-square method (WLS) to do a linear regression for the bin-averaged data.
\end{itemize}
Since the errors from numerical variations and stochastic modelling become zero when averaging the data within each bin, the simple linear regression for the bin-averaged data should yield an  unbiased estimate of the slope.

To confirm the effectiveness of this method, we use it to fit the observational data of massive galaxies and classical bulges reported in \citet{kormendy2013} and compare with other methods. We use the bisector method \citep{isobe90}, $\chi^2$ estimator method \citep{tremaine2002}, and {\rm the} grouping method introduced here to determine the slope and intrinsic scatter, with results shown in Table~\ref{tab:res}. {Since our simulated data lacks explicit measurement errors at this stage, this comparison is only performed on the observational dataset.}

\begin{table}
  \caption{The slope and scatter in the $M_{\mathrm{BH}} - M_{\star}$ relation,
    estimated for the observational data in \citet{kormendy2013}
    by either using the bisector method, the $\chi^2$ estimator method, or the grouping method.
    All three approaches show good consistency.}
 \centering
 \begin{tabular}{lccc}
  \hline
 Methods & Bisector & $\chi^2$ estimator & Group\\
 \hline
 Slope & 1.165 & 1.15 &  1.17\\
 Scatter & 0.2811 & 0.2809 & 0.2813\\
 \hline
 \end{tabular}
 \label{tab:res}
\end{table}

From the table, we can infer that the three methods obtain similar results. The estimate of the intrinsic scatter with the grouping method is slightly higher than for the other two methods. However, the relative difference is less than 0.1\%, demonstrating that the grouping method is a valid approach for estimating the slope and the scatter. We will thus use it below to fit the simulation data, allowing us to compare the intrinsic scatter of the observational data with the scatter measured for the cosmological simulation models.

\begin{figure}
   \includegraphics[width=0.45\textwidth]{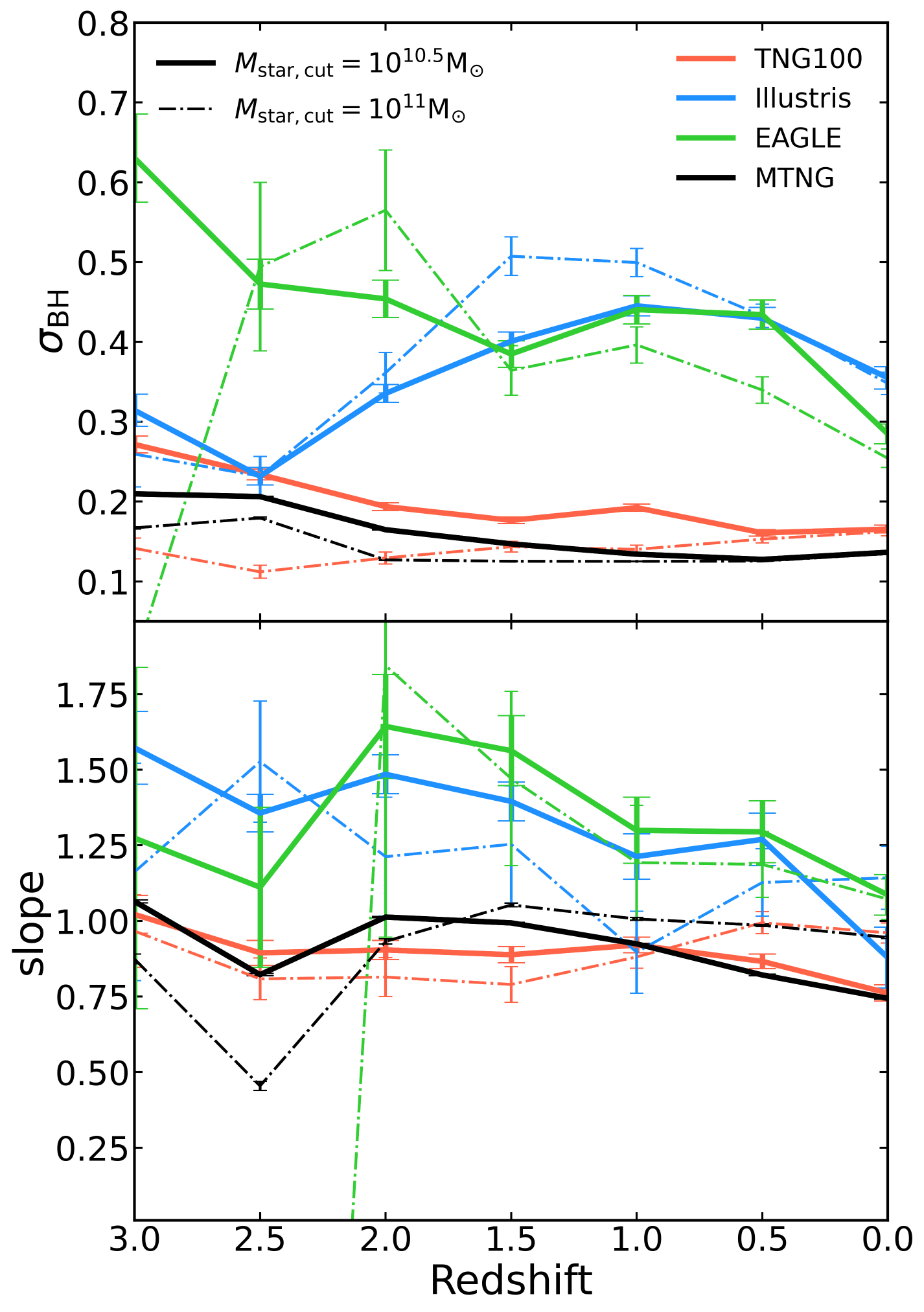}
   \caption{\textit{Upper panel:} The time evolution of scatter in the $M_{\mathrm{BH}} - M_{\star}$ relation in the TNG100, Illustris, EAGLE, and MTNG simulations, obtained by using the fitting procedure mentioned in Section~\ref{sec:fit}. \textit{Lower panel:} The time evolution of the corresponding slope in TNG100, Illustris, EAGLE, and MTNG. The solid lines represent the results obtained from the data samples with $M_{\star}>10^{10.5}\,{\rm M_{\odot}}$, while the dashed-dotted lines give the results obtained from data samples with the cut $M_{\star}>10^{11}\,{\rm M_{\odot}}$. These results show that the scatter and slope of the $M_{\mathrm{BH}} - M_{\star}$ relation in four simulations can be clearly separated into two groups, independent of the selection criteria.}
    \label{sigma&k}
\end{figure}

\begin{figure}
   \includegraphics[width=0.45\textwidth]{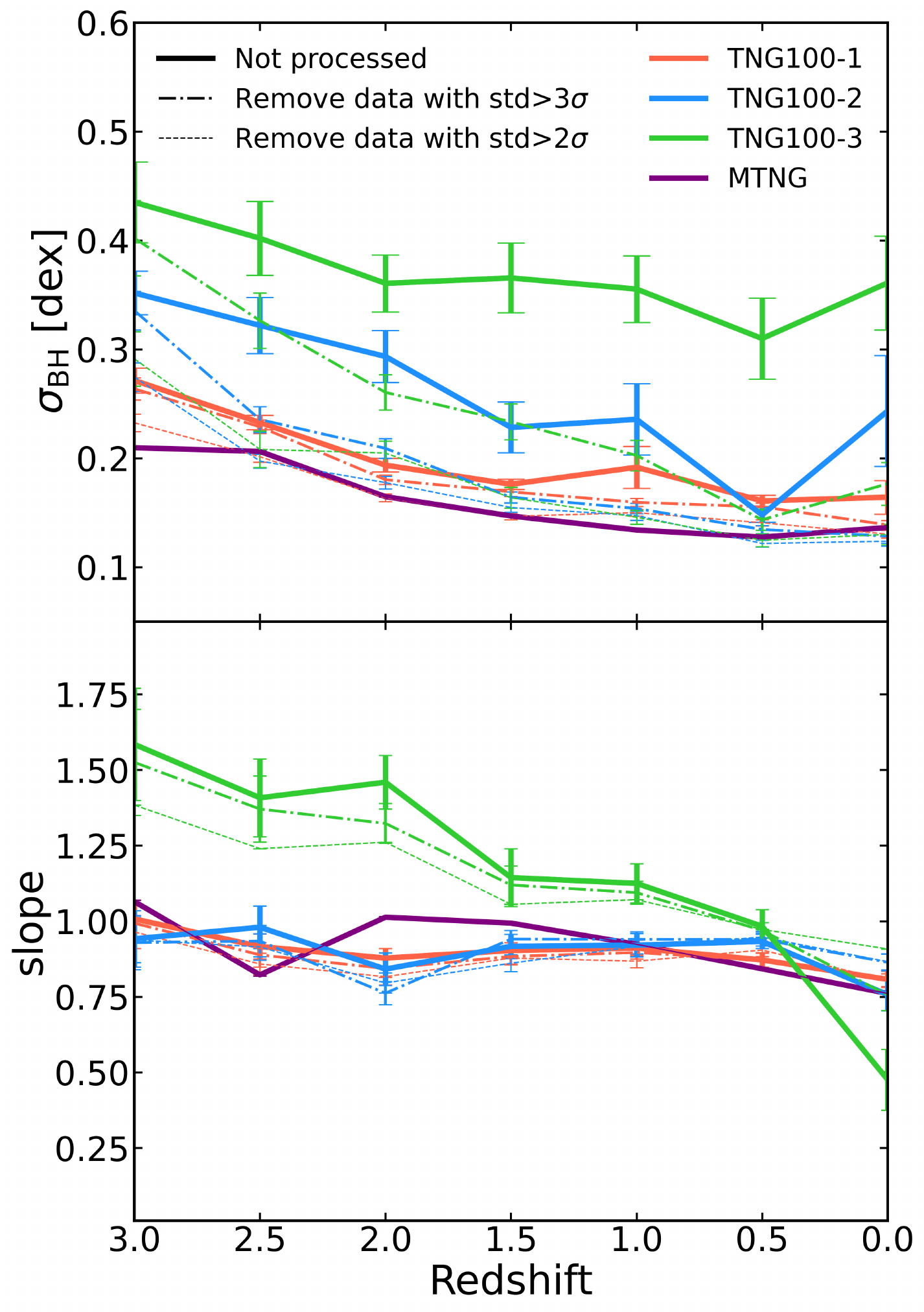}
   \caption{\textit{Upper panel:} The time evolution of scatter in the TNG100-1, TNG100-2, and TNG100-3 simulations, obtained by using the fitting procedure described in Section~\ref{sec:fit}. \textit{Lower panel:} The time evolution of the slope in the TNG100-1, TNG100-2, and TNG100-3 runs. Different line styles, as labeled, represent the results obtained without data cleaning, with removing data outside 2$\sigma$, and with removing data outside 3$\sigma$. The results show that the slope and scatter converge in TNG100-1, TNG100-2, and TNG100-3 after removing outliers.}
    \label{res_sigma&k}
\end{figure}

\subsection{Measured values of slope and scatter}

\subsubsection{Time evolution}

In this section, we investigate the time evolution of the scatter and slope of the $M_{\rm BH}- M_{\rm \star}$ relation from $z=3$ to $z=0$ in Illustris, TNG, EAGLE and MTNG. Figure~\ref{sigma&k} presents this time evolution for the data samples with $M_{\star}>10^{10.5}\,{\rm M_{\odot}}$ (solid lines) and $M_{\star}>10^{11}\,{\rm M_{\odot}}$ (dashed-dotted lines) obtained by using the fitting procedure described in Section~\ref{sec:fit}. It is noted that the fitting results for EAGLE at $z=3$ and $z=2.5$ with $M_{\star}>10^{11}\,{\rm M_{\odot}}$ formally have a zero or even negative slope, but this is simply due to too few data points for the corresponding samples. Therefore, we shall ignore the results from the data samples with $M_{\star}>10^{11}\,{\rm M_{\odot}}$ at $z\geq2.5$ for EAGLE.

The fitting results for the data with the $M_{\star}>10^{10.5}\,{\rm M_{\odot}}$ selection criterion at different redshifts and for different stellar mass cuts are also included in Figure~\ref{selecteddata}. For the samples with $M_{\star}>10^{10.5}\,{\rm M_{\odot}}$, it can be seen from the figures that the four simulations can be roughly divided into two groups. TNG100 and MTNG have similar values for scatter and slope, with the scatter in both simulations varying from $\sim$0.25 to $\sim$0.15 dex, and the slopes ranging from $\sim$1 to $\sim$0.75 between  $z=3$ and $z=0$. {Despite the significant resolution difference between the two simulations, this consistency suggests that the slope and scatter of the $M_{\rm BH}-M_\star$ relation are robust to resolution effects and reflect more fundamental aspects of BH–galaxy co-evolution.} In comparison, EAGLE and Illustris show higher scatter, and also a somewhat steeper slope. The values of the scatter in EAGLE vary from $\sim0.6$ to $\sim$0.3 dex from $z=3$ to $z=0$. In Illustris, the value of the scatter increases from $\sim 0.3$ to $\sim0.5$ dex over $z=3$ to $z=1$, and then decreases from $\sim 0.5$ to $\sim0.35$ dex over the range $z=1$ to $z=0$, {although the latter is not statistically significant based on a t-test}. With respect to the evolution of the slope, the values in Illustris and EAGLE tend to be higher than unity at high redshift, and then decrease to $\sim1$ by $z=0$.

Since the TNG100 and MTNG simulations share a very similar galaxy formation model with only the box size and the numerical resolution being different, the similar time evolution of scatter and slope indicates a good consistency of the models for the predicted time evolution of the $M_{\rm BH}-M_{\star}$ relation, which is also in line with the findings in \citet{parkmor2023}. Since galaxy mergers are expected to reduce the scatter and make the slope closer to 1 \citep{peng2007,hirschmann2010,Jahnke2011}, the results for TNG100 and MTNG indicate that mergers mainly contribute to the growth of BHs only at the massive end in TNG100 and MTNG. The Illustris and EAGLE simulations on the other hand show a larger scatter and steeper slope compared to TNG100 and MTNG, suggesting that star formation or BH accretion still significantly contribute to the growth of the massive galaxies in these simulations, or in other words, that they are less strongly quenched than corresponding galaxies in TNG100 and MTNG. (See also the discussion in Section~\ref{sec:dicussion} and Figure~\ref{sfr_bhar}.) 

Although we expect that the scatter is not influenced  by our estimation procedure of the slope, the intrinsic scatter is still hard to obtain from the simulation data since it may also have contributions from numerical variations and stochastic modelling, described by the last three terms on the right hand side of equation~(\ref{scatterformula}). To compare with the intrinsic scatter of observations, we thus need to estimate the numerical contributions to the scatter. We will therefore try to separate the contributions to the scatter coming from different sources, and present a detailed analysis of this point in Section~\ref{sec:origin}.

\subsubsection{Impacts of the selection criteria}

To investigate the impacts of the selection criteria on the fitting results for the scatter and slope, we compare the results obtained for data samples with $M_{\star}>10^{10.5}\,{\rm M_{\odot}}$ and $M_{\star}>10^{11}\,{\rm M_{\odot}}$. It can be seen from Fig.~\ref{sigma&k} that the results for the two samples with different stellar mass cuts are very similar. The slopes of the samples with $M_{\star}>10^{11}\,{\rm M_{\odot}}$ are a bit closer to 1 in all simulations. This can be understood as a consequence of the natural trend of the $M_{\rm BH}-M_{\star}$ relation to evolve towards a slope close to 1 when galaxies are quenched (which is mostly the case for massive galaxies in all four simulations) and mergers dominate the growth of BHs and galaxies \citep{peng2007}.  These results suggest that our selection criteria have a minor influence on the results obtained above.

\subsubsection{Impacts of the numerical resolution}\label{sec:res}

The numerical resolution may also affect the results obtained from the simulations. We use the TNG100 simulations as an example to analyze the possible impact of resolution on the scatter we measure. Figure~\ref{res_sigma&k} displays the time evolution of the scatter and slope of the data samples with our default selection criteria for the TNG100-1, TNG100-2, and TNG100-3 simulations. The dotted line gives the results of the data samples without any preprocessing. From these results we can infer that the scatter increases with decreasing numerical resolution. The measured slope is almost identical in TNG100-1 and TNG100-2, while the slope of the TNG100-3 sample is a bit larger than TNG100-1 and TNG100-2 at high redshift, but then decreases with decreasing redshift. At $z=0$, the slope measured for the three different resolutions of TNG100 is reassuringly similar. The higher slope in the low-resolution calculation likely arises from gas density estimates that are biased low in poorly resolved galaxies, causing an underestimate of the BH accretion rate especially in low-mass galaxies, steepening the resulting slope in the low-resolution simulation.

We also find that the BH masses of galaxies exhibit larger scatter in the low resolution simulations, with a tendency to also produce objects far away from the scaling relation. When we remove those data points that are outside a $2\sigma$ or $3\sigma$ region around the relation, where $\sigma$ is the standard deviation of the BH mass relative to the scaling relation, we obtain the dashed-dotted and solid lines in Figure~\ref{res_sigma&k}. For the scatter, we then find that the results become much closer again when we remove these outliers, especially after removing the points with deviations larger than $2\sigma$. For the slope, we find that the differences for the different treatments are generally small, suggesting that the outlier removal has little effect on the global trends. Based on these results, we conclude that low resolution in the TNG model can produce a population of galaxies far away from the $M_{\rm BH}-M_{\star}$ relation. This tail of objects is probably originating in discreteness noise. When they are filtered out, the results for scatter and slope obtained from the bulk of the galaxies are not strongly affected by numerical resolution. 

\begin{figure*}
   \includegraphics[width=0.85\textwidth]{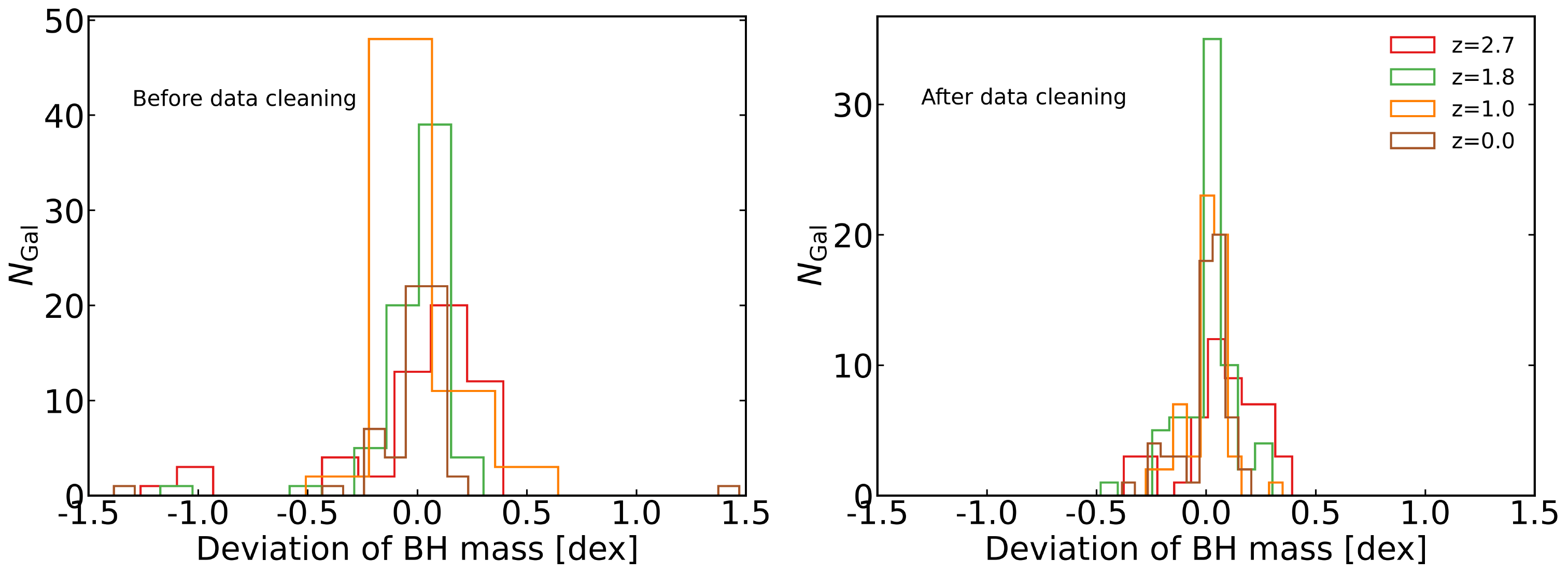}
   \caption{\textit{Left panel:} The standard deviation of the BH mass in the matched subhalos of five identical TNG simulations described in Section~\ref{sec:expriment} without any data cleaning. \textit{Right panel:} The standard deviation of the BH mass in the matched subhalos of five identical TNG simulations described in Section~\ref{sec:expriment}, but now with outlier removal of data outside 3$\sigma$. Through comparing these two figures we can see that after removing outliers, the bulk of the data points is consistent with a Gaussian distribution.}
    \label{sigClean}
\end{figure*}

\section{Origin of the scatter in cosmological simulations} \label{sec:origin}

In this section, we discuss possible sources of scatter in the $M_{\mathrm{BH}}-M_{\mathrm{\star}}$ relation in different cosmological simulations. We also carry out numerical experiments we have designed for identifying the magnitude of scatter from different origins.

\subsection{Possible sources of scatter}\label{origins}

In equation~(\ref{scatterformula}), the second and third terms on the right hand side disappear if the estimation of the slope is unbiased, then only two terms are left. The last term adds scatter due to numerical variations and stochastic modelling, which can be classified as being of purely numerical origin. The numerical sensitivities arise  due to the highly non-linear nature of the gravity-hydrodynamical processes in galaxy formation. Here small variations originating in minor differences from numerical round-off or force errors from the TreePM gravity solver can be amplified significantly, which is often referred to as chaotic behaviour or butterfly effects \citep{genel19,keller19,borrow23b}. Note also that due to computational limitations, it is impossible to model many astrophysical processes like star formation or BH accretion from first principles. Instead, these phenomena occur ``subgrid'', and are often modeled with the help of stochastic treatments. But introducing pseudo random numbers in cosmological simulations induces additional variations in the physical quantities inferred from the simulations, because individual resolution elements do not get to see the same random numbers any more when computations have already started to slightly diverge due to the more subtle reasons mentioned earlier. 

The first term in equation~(\ref{scatterformula})  represents the intrinsic scatter that comes from the astrophysical processes themselves. The origin of this intrinsic scatter can be broadly divided into two main processes -- hierarchical merging on one side, and BH accretion and associated feedback on the other side. The galaxies and their central BHs  experience a series of mergers across the cosmological evolution. Since the galaxies do not strictly lie on the scaling relation when they merge, the merger remnant realizes scatter for the galaxies. On the other hand, although the existence of a co-evolution of BHs and their host galaxies may regulate galaxy evolution, making galaxies evolve towards the $M_{\rm BH}-M_{\star}$ relation and reducing the scatter, the BH accretion in itself can also evolve BHs away from the relation and introduce scatter in the scaling relation. The degree to which this happens depends on the physics model for galaxy evolution, and in particular on the feedback for BH growth and star formation. When feedback self-regulates the BH growth, the combination of accretion and feedback can also reduce scatter. Thus, while accretion is a driver of scatter, feedback can in fact act in the opposite direction by reducing it through self-regulation, particularly in accretion-dominated regimes. Aside from these physical processes, another important source for scatter in the $M_{\rm BH}-M_{\star}$ relation can lie in variations of the BH seed mass, especially when these seeds are comparatively massive so that substantial galaxy growth is needed before all memory of the seed variations is lost.

To summarize this section, we can consider a simple model to describe the origin of the variances of the BH mass in the $M_{\rm BH}-M_{\star}$ relation:
\begin{equation}
\begin{aligned}
\sigma_{\rm BH}^2
&=\underbrace{\sigma_{\rm BH\,seed}^2 + \sigma_{\rm Acc\&FB}^2
+\sigma_{\rm HM}^2}_{\rm physical}\\
&+\underbrace{\sigma_{\rm SM,\, BH}^2+\sigma_{\rm Numerics,\,BH}^2 + \beta^2\sigma_{\rm SM,\, \star}^2+\beta^2\sigma_{\rm Numerics,\,\star}^2}_{\rm numerical}\\
&+\sigma_{\rm oths}^2 +\sigma_{\rm cross}^2 .
\end{aligned}
\label{scattermodel}
\end{equation}
Here,
\begin{itemize}
    \item $\sigma_{\rm BH\,seed}^2$ is the variance coming from the variation of BH seed mass;
    \item $\sigma_{\rm Acc\&FB}^2$ is the variance due to the variation of BH accretion and feedback processes;
    \item $\sigma_{\rm HM}^2$ is the variance introduced by variations of the merging history of different halos;
    \item $\sigma_{\rm SM,\, BH}^2$ is the variance from stochastic modelling of BHs in simulations;
    \item $\sigma_{\rm Numerics,\,BH}^2$ is the variance from non-linear chaotic behaviour of the cosmological simulations;
    \item $\beta^2\sigma_{\rm SM,\, \star}^2+\beta^2\sigma_{\rm Numerics,\,\star}^2$ are the corresponding terms for the modelling of the stellar mass, which indirectly enter the inferred variance of the BH mass in the $M_{\rm BH}-M_{\star}$ relation;
    \item $\sigma_{\rm oths}^2$ is the variance from other processes we do not consider explicitly in this model;
    \item and $\sigma_{\rm cross}^2$ finally captures cross-covariances between terms.
\end{itemize}
In this framework, the first three contributions are physical in origin, whereas the subsequent four are numerical. 
For simplicity, we ignore cross terms in the following analysis.

\subsection{Numerical experiments}\label{sec:expriment}

To better characterize different sources of scatter in the $M_{\rm BH}-M_{\star}$ relation, we carry out different numerical experiments, using the TNG model as a fiducial test case. We begin by estimating the scatter from numerical variations and stochastic modelling. To this end we run five TNG box simulations with a box size of $25\,h^{-1} {\rm cMpc}$, using identical initial condition but different numbers of compute nodes. The latter is a means to introduce  implicit numerical variations because the {\small AREPO} code is only binary invariant in its results if the number of compute nodes/cores is kept unchanged for a repeated simulation. Varying this will change the domain decomposition, and thus the exact order in which gravitational forces and hydrodynamical fluxes are added up, both of which lead to floating point round-off variations. In addition, the random numbers that will be applied to individual gaseous resolution elements for the stochastic modelling of star formation will be completely different when the number of cores for parallel execution is changed.

For the five results we then perform an object-by-object subhalo matching and calculate the variance and standard deviation for the sample of matched subhalos. To match subhaloes between simulations, we adopt a straightforward strategy based on spatial proximity and subhalo mass. Since all simulations are initialized with identical conditions and differ only in the number of computational nodes, the large-scale structures evolve nearly identically. For each central galaxy, we search for nearest counterparts within 30 ckpc in three-dimensional space and require their stellar masses to agree within 10\%. We restrict the matching to central galaxies (excluding satellites) to ensure robustness. With this method, we achieve a matching fraction of approximately 60–70\% across the full central galaxy population, increasing to ~80\% for the massive galaxies that are the primary focus of this study. For each matched set, we calculate the standard deviation of the quantity of interest (e.g. BH mass) among the five simulations. This provides an estimate of the numerical variation for that particular subhalo. To obtain a representative scatter for a given stellar mass, we group the matched subhaloes into stellar mass bins and report the median of the individual standard deviations within each bin. Since the initial conditions are identical, the variation in galaxy properties for the matched subhalos can be attributed to numerical variations and stochastic modelling. Note that \citet{genel19} reported that the variations in galaxy properties measured in TNG box simulations through a similar method is not sensitive to the employed numerical resolution.

\begin{figure}
   \includegraphics[width=0.45\textwidth]{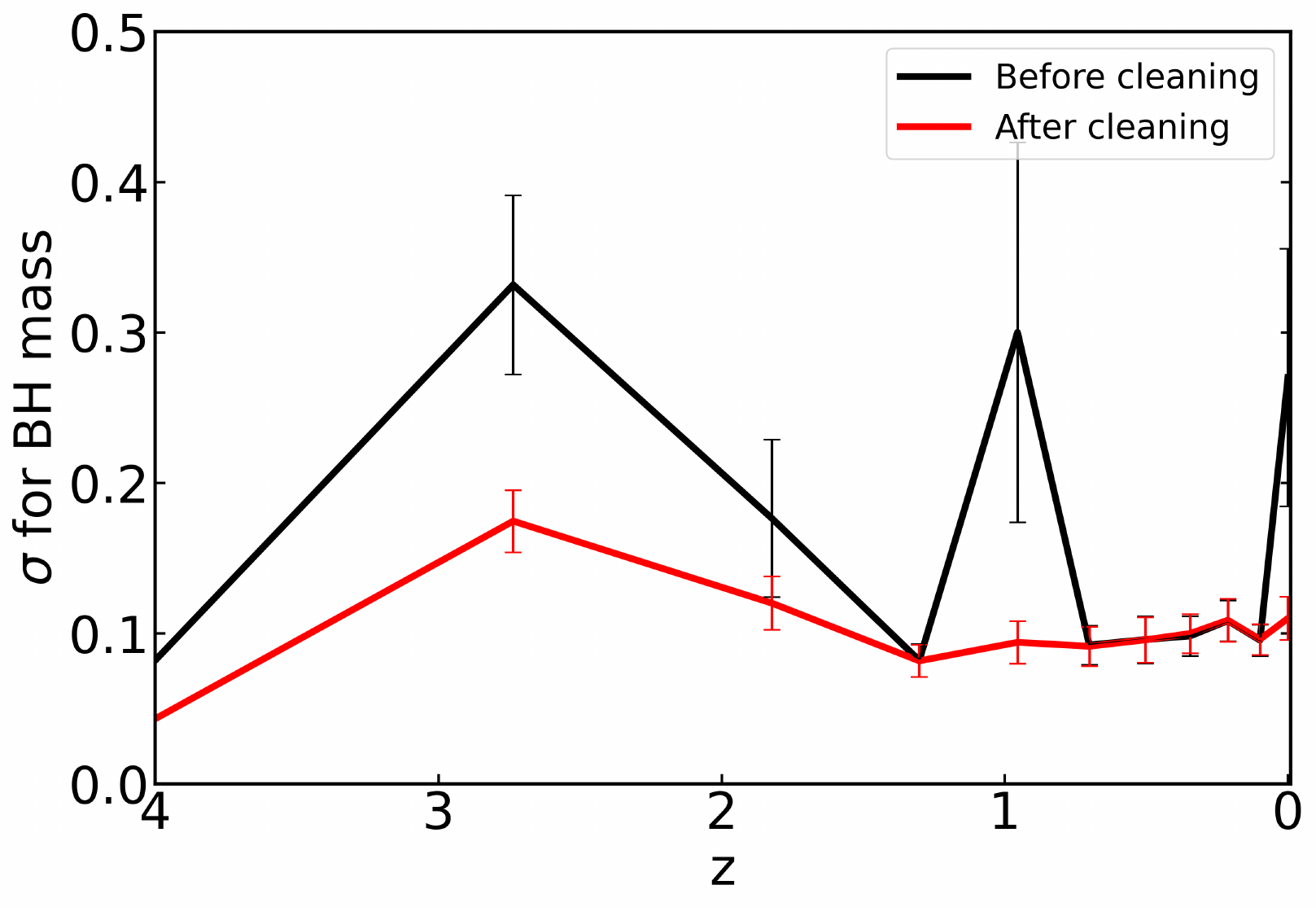}
   \caption{The time evolution of the variance of the BH mass in  matched halos of five identical TNG simulation with and without data cleaning. The figure indicates that after removing data outliers the scatter in the BH mass from numerical origins shows a time-independent behaviour.}
    \label{BHnumscev}
\end{figure}

The left panel of Figure~\ref{sigClean} shows the standard deviation of the BH mass in our five TNG simulation with identical initial condition, as mentioned above. The galaxies are also fulfilling the selection criteria mentioned in Section~\ref{sec:selectcriteria}. We  find that the variations of the BH mass roughly follow a Gaussian distribution, except that there are also some data points far away from zero. Previous work \citep[e.g.,][]{genel19} has shown that the stochastic numerical scatter between simulations with identical initial conditions is well described by a Gaussian distribution, and that the rare extreme deviations may be typically caused by technical issues such as matching failures. Including these outliers would substantially overestimate the variance of the BH mass and obscure the underlying scatter. Since our aim is to quantify the typical numerical fluctuations rather than the full statistical spread including rare anomalies, we exclude data points outside a $3\sigma$ region in the analysis. As shown in the right panel of Figure~\ref{sigClean}, the distribution of the data is well preserved when these outliers are removed. Figure~\ref{BHnumscev} compares the time evolution of the measured variance of the BH mass before and after cleaning of the data as mentioned above. When outliers are removed, the variance shows a well behaved behaviour with a value that is roughly constant across $z=3$ to $z=0$, which is consistent with the findings in \citet{genel19}. We will thus adopt this data cleaning strategy for estimating the size of the scatter due to numerical noise, denoted $\sigma^2_{\rm BH}$ in equation~(\ref{scatterformula}). Subtracting this from the measured full scatter  $\hat\epsilon^2$, we can then also obtain an estimate of the intrinsic physical scatter $\epsilon^2$, which can be viewed as being composed of two main sources, hierarchical merging and gas accretion.

We have additionally run five further TNG box simulations using a similar treatment as mentioned above, but turning the BH accretion off.  For these simulations, variations due to BH accretion are not introduced by construction, such that the scatter in these simulations is only caused by hierarchical merging and numerical variations. Assuming that the scatter from different sources is approximately uncorrelated and additive, we can estimate the scatter just from BH accretion by subtracting the corresponding result from the full scatter obtained earlier. Here we for simplicity ignore the stellar mass dependence of the scatter, so this is evidently only a coarse estimate. By using this approach, we can separately estimate the contributions to the intrinsic scatter from hierarchical merging and gas accretion. 

\begin{table*}
  \caption{Description of the numerical experiments designed to isolate different components of the scatter model (Equation~(\ref{scatterformula})).}
 \centering
 \begin{tabular}{ll}
  \hline
 simulations & descriptions\\
 \hline
 TNGFid & a 25 Mpc/$h$ box simulation with $2\times256^3$ resolution elements, using the fiducial TNG model\\
 TNGSeries & five TNGFid simulations with identical setup but different number of nodes\\
 NoAcc & The same as TNGFid simulations but turning of BH accretion\\
 NoAccSeries & five NoAcc simulations with identical setup but different number of nodes\\
 \hline
 \end{tabular}
 \label{tab:des}
\end{table*}

\begin{table*}
  \caption{Measured variances of BH mass from the different numerical experiments, corresponding to the terms in the scatter model (Equation~(\ref{scatterformula})).}
 \centering
 \begin{tabular}{ll}
  \hline
 simulations & descriptions\\
 \hline
 TNGFid & $\sigma_{\rm Acc\&FB}^2+\sigma_{\rm HM}^2+\sigma_{\rm SM,\, BH}^2+\sigma_{\rm Numerics,\,BH}^2 + \beta^2\sigma_{\rm SM,\, \star}^2+\beta^2\sigma_{\rm Numerics,\,\star}^2+\sigma_{\rm oths}^2$\\
 TNGSeries & $\sigma_{\rm SM,\, BH}^2+\sigma_{\rm Numerics,\,BH}^2 + \beta^2\sigma_{\rm SM,\, \star}^2+\beta^2\sigma_{\rm Numerics,\,\star}^2+\sigma_{\rm oths}^2$\\
 NoAcc & $\sigma_{\rm HM}^2+\sigma_{\rm Numerics,\,BH}^2 + \beta^2\sigma_{\rm SM,\, \star}^2+\beta^2\sigma_{\rm Numerics,\,\star}^2+\sigma_{\rm oths}^2$\\
 NoAccSeries & $\sigma_{\rm Numerics,\,BH}^2 + \beta^2\sigma_{\rm SM,\, \star}^2+\beta^2\sigma_{\rm Numerics,\,\star}^2+\sigma_{\rm oths}^2$\\
 \hline
 \end{tabular}
 \label{tab:var}
\end{table*}

As described above, each experiment isolates a different subset of the variance terms in our scatter model (Equation~\ref{scatterformula}). We summarize the design of the experiments and the corresponding variances in Tables~\ref{tab:des} and \ref{tab:var}. Linear combinations of these results then allow us to estimate the individual contributions of the components $\sigma_{\rm Acc\&FB}^2$, $\sigma_{\rm HM}^2$, $\sigma_{\rm SM,\, BH}^2$, $\sigma_{\rm Numerics,\, BH}^2$, $\sigma_{\rm SM,\, \star}^2$, and $\sigma_{\rm Numerics,\, \star}^2$ in the model.

\subsection{The scatter in TNG from different origins}\label{sec:diff_sc}

\begin{figure}
   \includegraphics[width=0.45\textwidth]{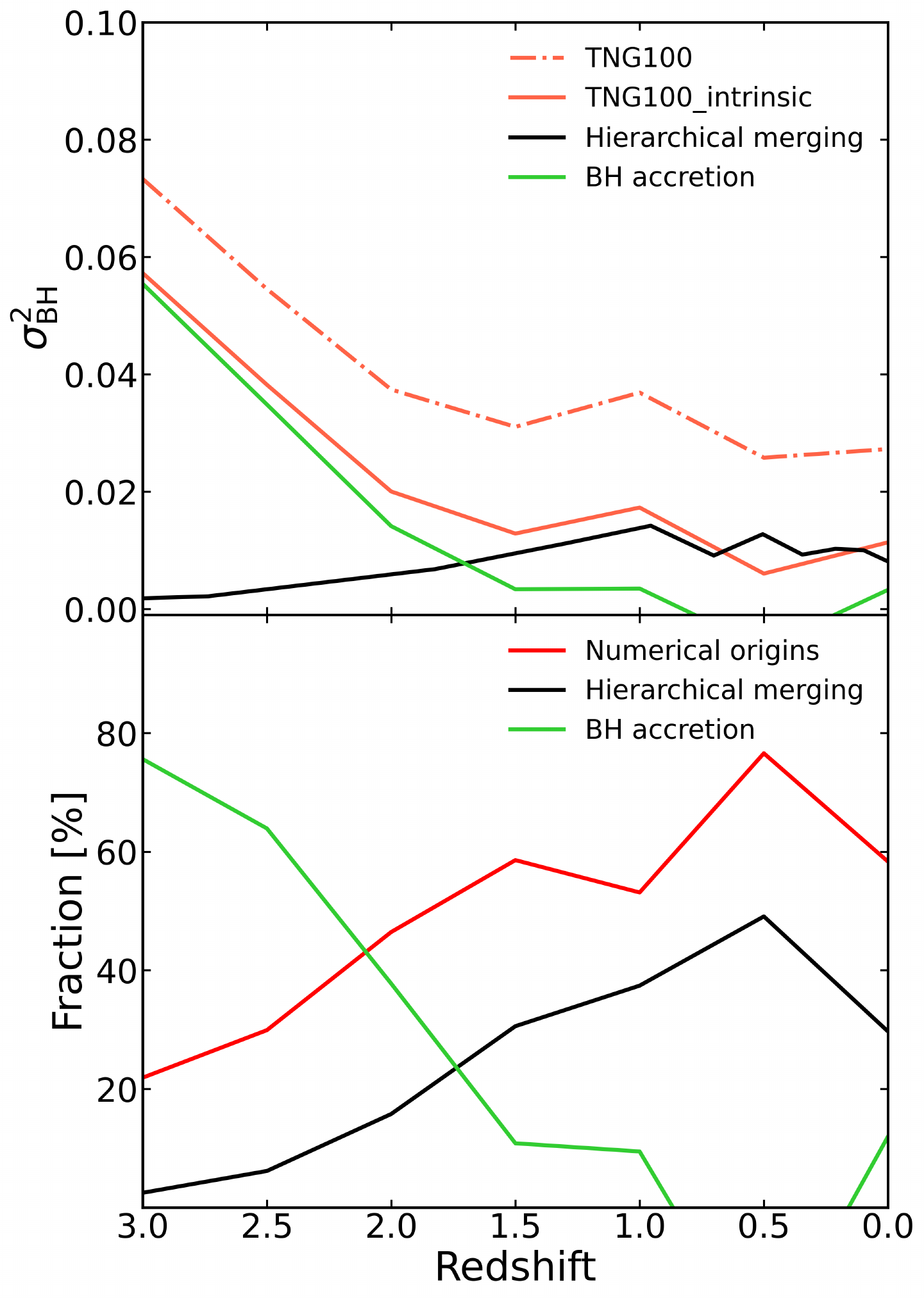}
   \caption{\textit{Upper panel:} The time evolution of the scatter of the $M_{\rm BH}-M_{\star}$ relation, the intrinsic scatter after correcting for the scatter from numerical origins, the scatter due to hierarchical merging, and {the} scatter due to BH accretion in the TNG100 simulation. \textit{Bottom panel:} The time evolution of fractional contributions to the scatter due to  numerical origins, hierarchical merging, and BH accretion in the TNG100 simulation. The statistical uncertainty of the results is $\sim10^{-5}$. The figure indicates that the scatter from numerical origins increasingly dominates from $z=3$ to $z=0$, reaching $\sim$60\% at $z=0$, whereas the scatter from accretion becomes less significant with time. At $z=0$, the scatter from physical origins is primarily contributed by hierarchical merging.}
    \label{TNGscComponent}
\end{figure}

Based on the description in Section~\ref{origins}, the sources of the scatter in the $M_{\rm BH}-M_{\star}$ relation can be broadly divided into numerical and physical ones. As described in the previous section, with the help of two sets of toy simulations we can obtain estimates for the numerical contribution to the scatter, and separate the intrinsic scatter into contributions from hierarchical merging and BH growth by gas accretion. These results are summarized in the upper panel of Figure~\ref{TNGscComponent}. Note that all of the results here are obtained from galaxies fulfilling the selection criteria discussed in Section~\ref{sec:selectcriteria}. The dash-dotted and solid red lines show the scatter and intrinsic scatter in TNG100 cosmological simulations after subtracting the scatter from numerical origins. The difference between scatter and intrinsic scatter is $\sim$ 0.02, and our estimates for $\beta$, $\sigma_{\rm BH}$, and $\sigma_{\star}$ are $\sim$ 1, 0.01 and 0.01, respectively. The physical sources of scatter from hierarchical merging and BH accretion are also shown in the upper panel of Figure~\ref{TNGscComponent}. Note that at $z\sim0.5$ the intrinsic scatter is less than the scatter due to hierarchical merging. This is because the scatter from different origins is obtained by independent numerical experiments that estimate approximate values for different terms in the simple model mentioned in Sect.~\ref{origins}, a procedure that may cause slight discrepancies in the final result due to the involved uncertainties.

Note that the scatter from hierarchical merging is relatively low and increases with decreasing redshift, which arises because there massive galaxies increasingly merge with small galaxies that do not strictly lie on the $M_{\rm BH}-M_{\star}$ relation. In contrast, the inferred scatter from BH accretion decreases with decreasing redshift, and becomes consistent with zero at $z\lesssim0.5$.  This can be easily understood since the massive galaxies will tend to quench with decreasing redshift due to the AGN feedback, so that the contribution of BH accretion to the scatter scatter becomes lower and lower with time for massive galaxies and disappears at low redshift. We note that in realistic scenarios, mergers can indirectly enhance accretion by increasing $M_{\rm BH}$, which boosts the subsequent growth due to the $\dot{M}_{\rm BH} \propto M_{\rm BH}^2$ scaling. While this coupling is physically relevant, our approach treats such post-merger accretion as part of the accretion-driven component in order to maintain a clear separation between the direct contributions of hierarchical merging and black hole accretion.

We also display the time evolution of the fractional contributions of the scatter from different origins normalized to the total scatter in the bottom panel of Figure~\ref{TNGscComponent}. Note that the sum of the fractions is not strictly guaranteed to yield $100\%$ because the scatter from different origins has been obtained through different methods.  In addition, the scatter from different origins is not strictly independent. Nevertheless, the sum  of the fractions is close to 1, indicating that our estimates for the magnitude of the scatter from different sources are reasonably consistent despite being obtained from different methods, and that likewise correlations between the scatter from different origins are relatively weak. 

From the figure, it can also be seen that the scatter from the BH accretion decreases from a dominating $80\%$ at $z=3$ to something consistent with $\sim0\%$ at $z\sim1$. In contrast, the relative contribution of hierarchical merging increases from being negligible at $z=3$ to $\sim 40\%$ at $z\sim0$, indicating that hierarchical merging contributes significantly to the scatter of the $M_{\rm BH}-M_{\star}$ relation at $z=0$ in the {TNG100} simulations. It is however interesting to note that numerical sources are dominating the scatter at $z<2$ in the TNG model, therefore it would be misleading to directly compare the scatter obtained from the simulations with observations without  a corresponding correction. In fact, the scatter seen in {TNG100} at low redshift would be noticeable lower in the absence of numerical noise.

\begin{figure}
   \includegraphics[width=0.45\textwidth]{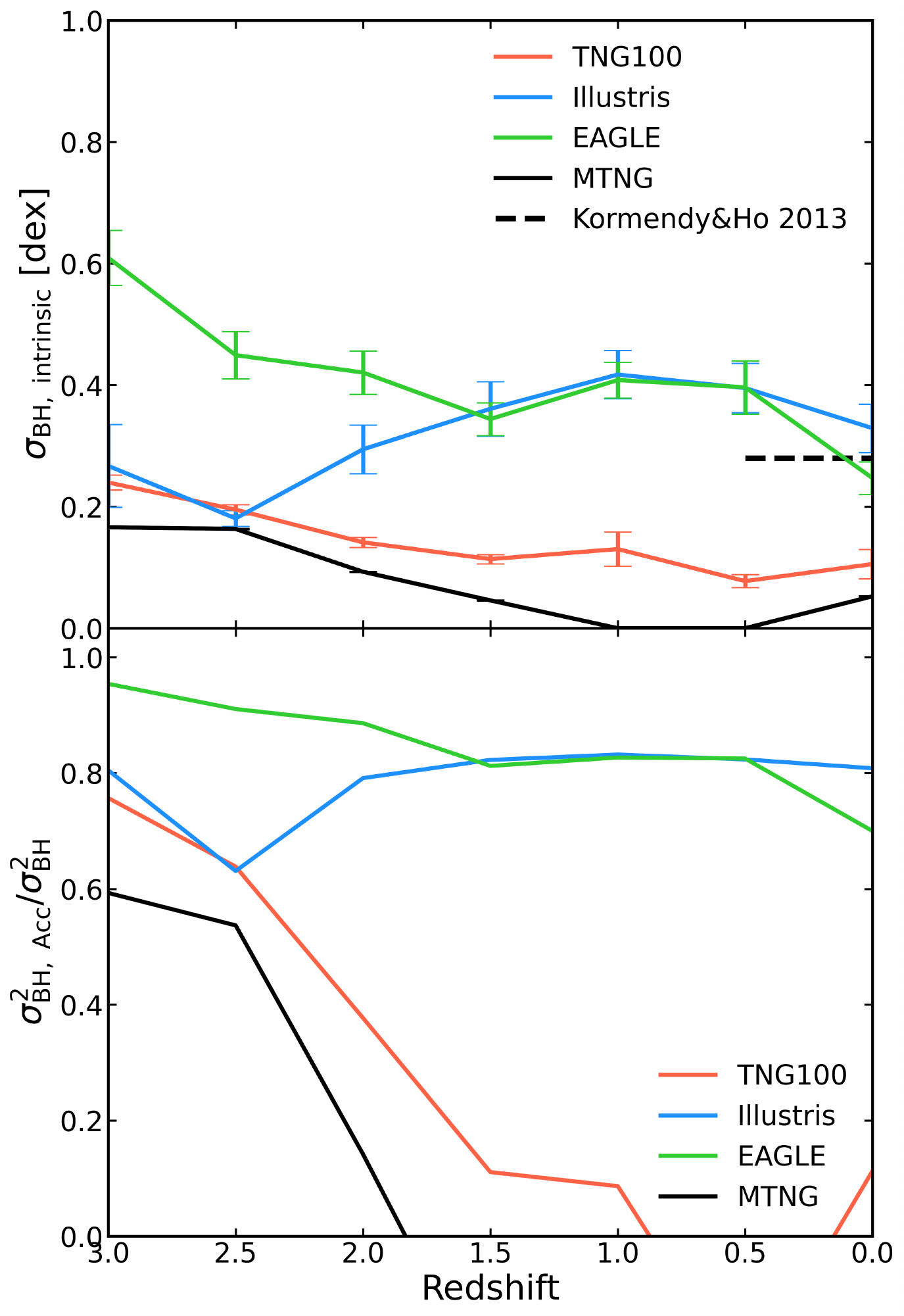}
   \caption{\textit{Upper panel:} The time evolution of the  intrinsic scatter of the $M_{\rm BH}-M_{\star}$ relation in TNG100, Illustris, EAGLE, and MTNG. \textit{Bottom panel:} The time evolution of the fraction of the scatter due to BH accretion with respect to the total scatter in the TNG100, Illustris, EAGLE, and MTNG simulations, based on the assumption that the scatter from hierarchical merging is similar in all simulations. {For the fraction of scatter from accretion, the four simulations show similar values at high redshift, but at $z=0$, a clear difference emerges, with MTNG and TNG showing a significantly lower proportion of scatter from accretion. This suggests that at low redshift the growth of black holes in MTNG and TNG is primarily driven by mergers.}}
    \label{Sc_intrinsic}
\end{figure}

\begin{figure}
   \subfigure{\includegraphics[width=0.46\textwidth]{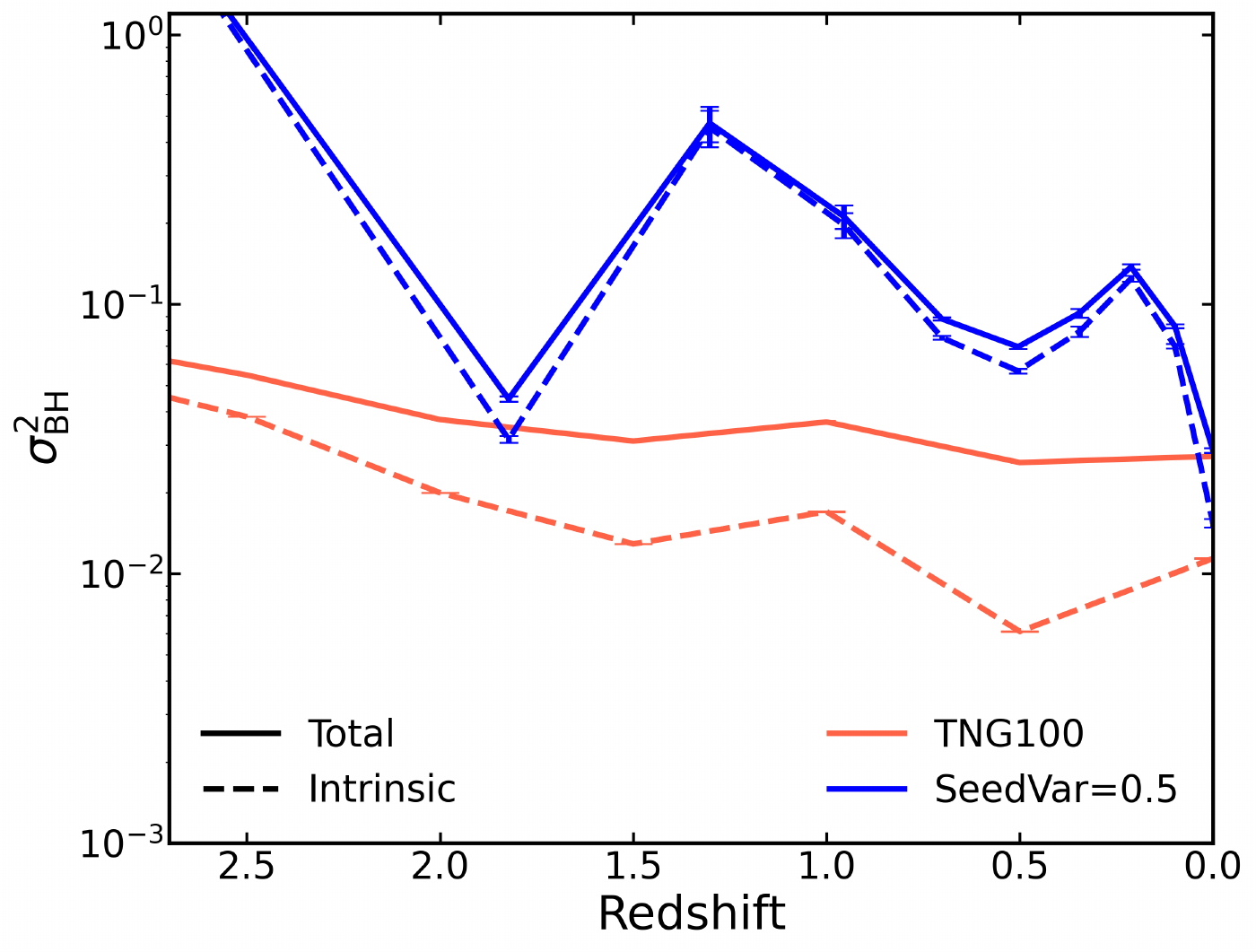}}\vspace{-3mm}
   \subfigure{\includegraphics[width=0.45\textwidth]{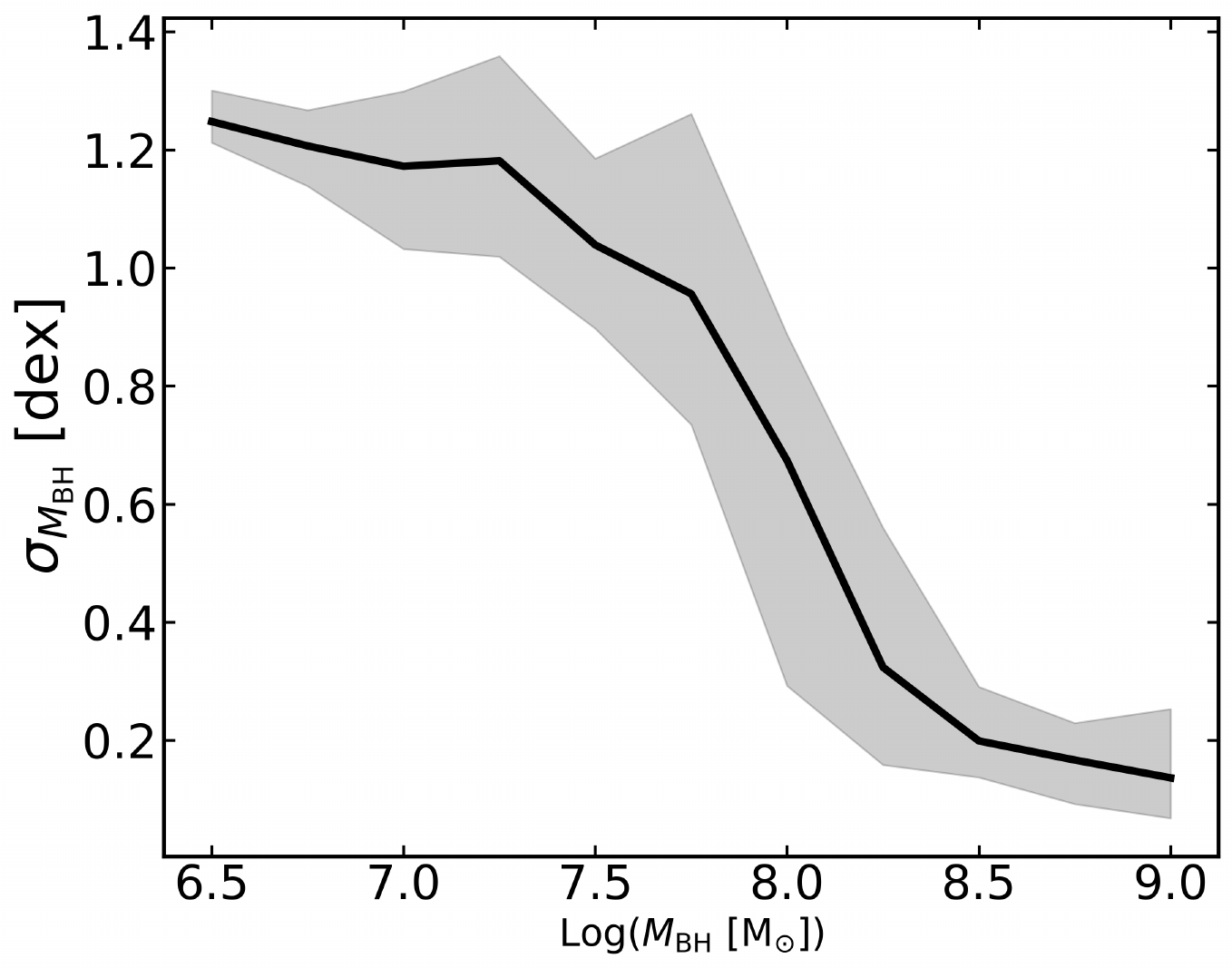}}\vspace{-3mm}
   \subfigure{\includegraphics[width=0.46\textwidth]{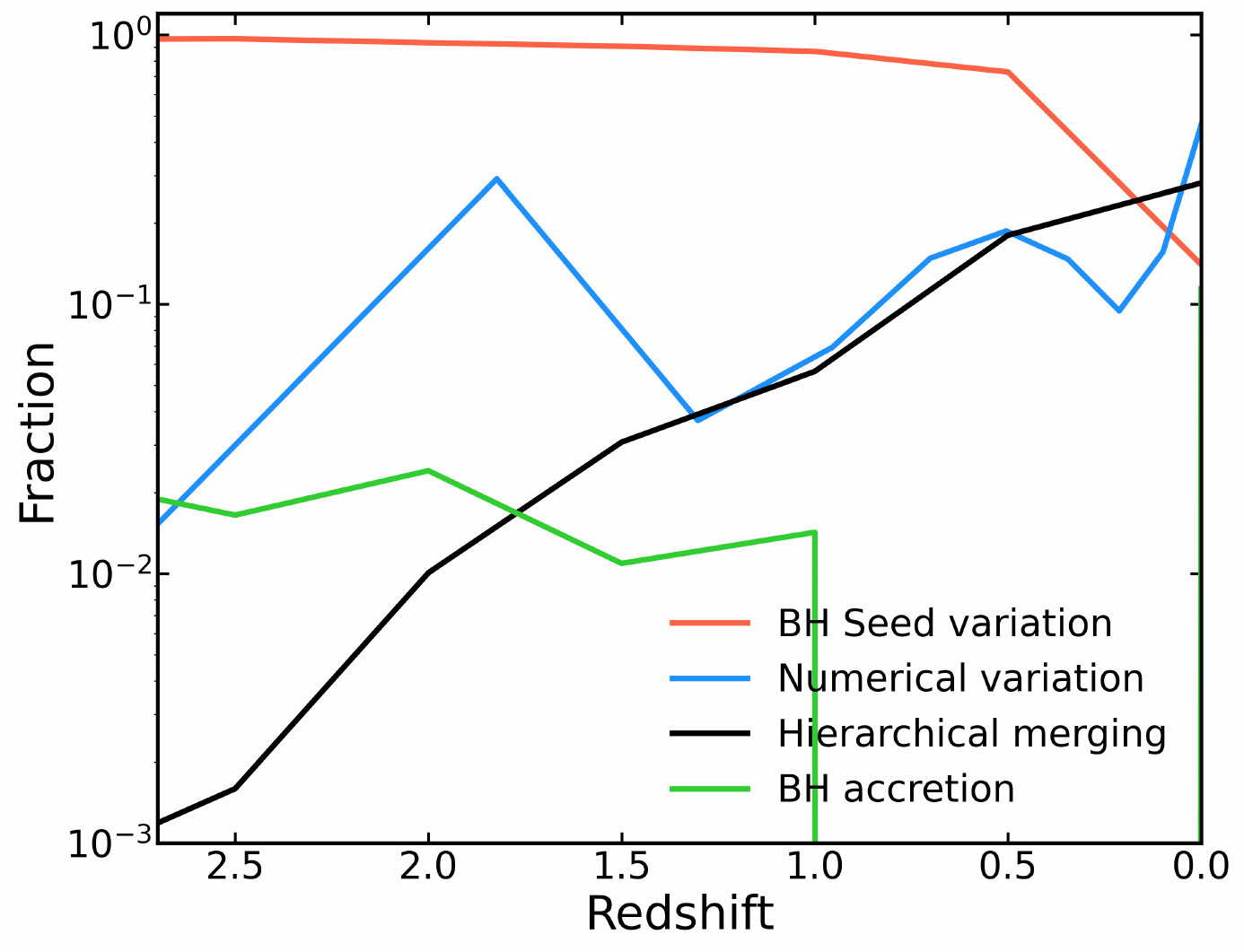}}
   \caption{\textit{Top panel:} The time evolution of the total and intrinsic scatter in the $M_{\rm BH}-M_{\star}$ relation in the TNG100 and SeedVar simulations. The ``SeedVar=0.5'' simulation imposes a random scatter of 0.5 dex on the BH seed mass in a small TNG-like simulation with box size 25 $h^{-1}$ cMpc and $2\times 256^3$ particles. 
   \textit{Middle panel:} {Scatter of BH mass across different mass bins in the SeedVar=0.5 simulation, stacked from $z = 3$ to $z = 0$. The shaded grey region denotes the 10-th and 90-th percentiles.} \textit{Bottom panel:} The time evolution of fractional contributions of different sources of scatter in the BH masses at different redshifts in SeedVar=0.5 simulation. The figure shows that the memory of an initial scatter in the BH seeds will be gradually erased and eventually becomes negligible in the TNG model due to the effective scatter reduction through mergers. We use a logarithmic $y$-axis scaling in the top and bottom panels. Since the values fall below $10^{-1}$ in some regions, this makes the differences between the lines easier to identify.}
    \label{var_results}
\end{figure}

\subsection{Other cosmological simulations and observational comparison}

We now try to apply the lessons learned from TNG about the different sources of scatter in the $M_{\rm BH}-M_{\star}$ relation to the Illustris, EAGLE, and MTNG runs. Instead of carrying out similar numerical experiments as described in Section~\ref{sec:expriment}, we simply assume that the scatter due to hierarchical merging is similar in these simulations compared to TNG. This should be a particularly good assumption for the MTNG simulation, which shares the simulation code and an almost identical galaxy formation model with TNG100. Although we do not explicitly compare the BH merger rates between MTNG and TNG, the expected differences in merger-driven scatter due to cosmic variance or volume effects should be small, given their shared cosmological model and the statistically representative population of massive halos in both volumes. It is somewhat less obvious that this assumption applies to the Illustris and EAGLE simulations. However, their BH seeding scheme is conceptionally very similar to TNG, suggesting that the time evolution of the scatter due to hierarchical merging should be broadly similar.

However, because the BH feedback and galaxy formation models in Illustris and EAGLE are quite different from TNG, the scatter due to numerical sources may be different. Thus we cannot directly apply the numerical scatter we obtained from TNG to the EAGLE and Illustris simulation data. Fortunately, \citet{borrow23b} measured the scatter due to the numerical variations in the EAGLE simulation. Based on their work, the variance of the BH mass $\sigma_{\rm BH}^2$ due to numerical origins in massive galaxies is around $0.01-0.03$, and for the stellar mass the scatter $\sigma_{\star}^2$ in massive galaxies is around 0.01 from numerical origins, which is comparable to what we inferred for TNG. Unfortunately, for Illustris there is no work in the literature that has reported the scatter from numerical origins. As Illustris shares many aspects of the galaxy formation model with TNG, and also uses the same code, we however expect these two simulations to be similar in practice with respect to the numerical scatter, so we shall assume the TNG numerical scatter also for Illustris.

Based on above assumptions, we can obtain the intrinsic scatter and the fractional contributions of different sources of scatter in these four cosmological simulations. We show the results in Figure~\ref{Sc_intrinsic}, where the upper panel displays the intrinsic scatter in the four simulations, while the black dash line at $z\sim 0$ is the intrinsic scatter obtained by \citet{kormendy2013}, who report a value of $\sim 0.29$ dex. This observational value is derived after correcting for typical measurement errors in BH mass estimates of $\sim0.12$ dex. It can be seen that TNG100 and MTNG have very similar intrinsic scatter, as expected, which amounts to $\sim0.1$ dex at $z\sim0$. Note that the intrinsic scatter can appear slightly negative in MTNG in some cases. This arises because the numerical contribution is not measured within MTNG itself, but instead approximated using results from our controlled numerical experiments as mentioned before, which may not perfectly align with the simulation-specific noise properties. In contrast, Illustris and EAGLE show similar scatter, with values comparable to the observations. Since hierarchical merging  reduces the scatter \citep{peng2007, hirschmann2010, Jahnke2011}, and a tighter co-evolution of BHs and their host galaxies also regulates systems to be close to the $M_{\rm BH}-M_{\star}$ relation \citep{kormendy2013,heckman2014,sun2015,zhuang2023}, we conclude that the low level of scatter in TNG100 and MTNG reflects a more efficient quenching of massive galaxies and  a particularly tight self-regulation of the $M_{\rm BH}-M_{\star}$ relation in this model. 

The bottom panel of Figure~\ref{Sc_intrinsic} shows the fractional contribution of the variance from BH accretion to the total variance. This contribution by BH accretion is $>50\%$ for TNG100 and MTNG at $z=3$ and then becomes negligible at low redshift. In contrast, in Illustris and EAGLE, the fraction of the variance from BH accretion is $\sim 80\%$ across the redshift range $z=3$ to $z=0$. This result indicates that the $M_{\rm BH}-M_{\star}$ relation is mainly regulated by hierarchical merging in TNG100 and MTNG at low redshift, with BH accretion becoming subdominant there, whereas Illustris and EAGLE appear to be less strongly quenched, so that variations in the BH mass by gas accretion are still more important. We will further discuss this issue in Section~\ref{sec:RegOrCoev}.

\subsection{Scatter due to BH seed mass variations}

As mentioned in Section~\ref{origins}, while the BH seed mass is fixed in the simulations, in the real Universe it may plausibly exhibit substantial variations, which could in turn affect the scatter in the $M_{\rm BH}-M_{\star}$ relation. The seed mechanism itself is presently unknown, and many different possibilities are actively discussed, including the run-away collapse of dense star clusters, the collapse of massive gas clouds, or the remnants of unusually massive pop-III stars. While the latter would still be comparatively light, the former may perhaps produce seeds with masses in excess of $10^5\,{\rm M}_\odot$. The seed masses could exhibit substantial variations \citep[e.g.][]{Bhowmick2024b}, and especially if they are relatively massive the initial spread may still affect the scatter of the BH masses later on.

BH seeding is implemented in the cosmological simulations we analyse in the present work simply by putting  BH seeds with a certain fixed mass into DM halos once they reach a prescribed threshold mass. Such a BH seeding implementation does not consider the environment, even though this is expected to influence the BH seed mass and the halo mass in which they are expected to form \citep{klessen2023}. Also, using a fixed BH seed mass minimizes -- by construction -- the  influence BH seed mass variations could have on the scatter of the $M_{\rm BH}-M_{\star}$ relation.

Providing a more elaborate physical model for BH seeding, such as those based on local gas conditions and dynamical criteria \citep[e.g.][]{Tremmel2017}, is beyond the scope of the current work. Instead we carry out a numerical experiment where we investigate the impact of varying the BH seed mass by running an extra cosmological simulation where we impose a random scatter of $0.5$ dex on the BH seed mass introduced in a small TNG-like simulation (hereafter SeedVar) with box size $25\,h^{-1} {\rm cMpc}$ and $2\times 256^3$ resolution elements.

\begin{figure*}
   \subfigure{\includegraphics[width=0.45\textwidth]{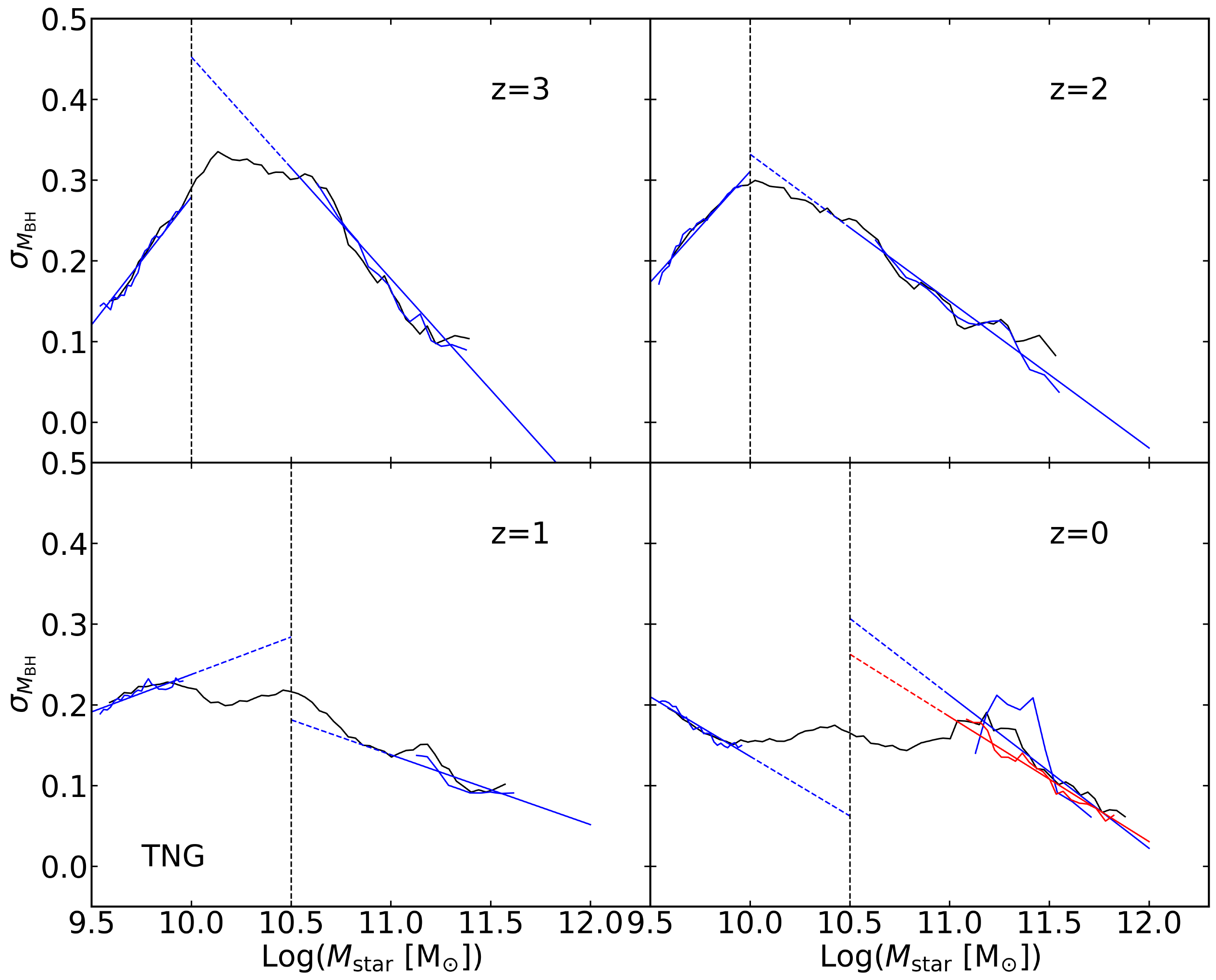}}
   \subfigure{\includegraphics[width=0.45\textwidth]{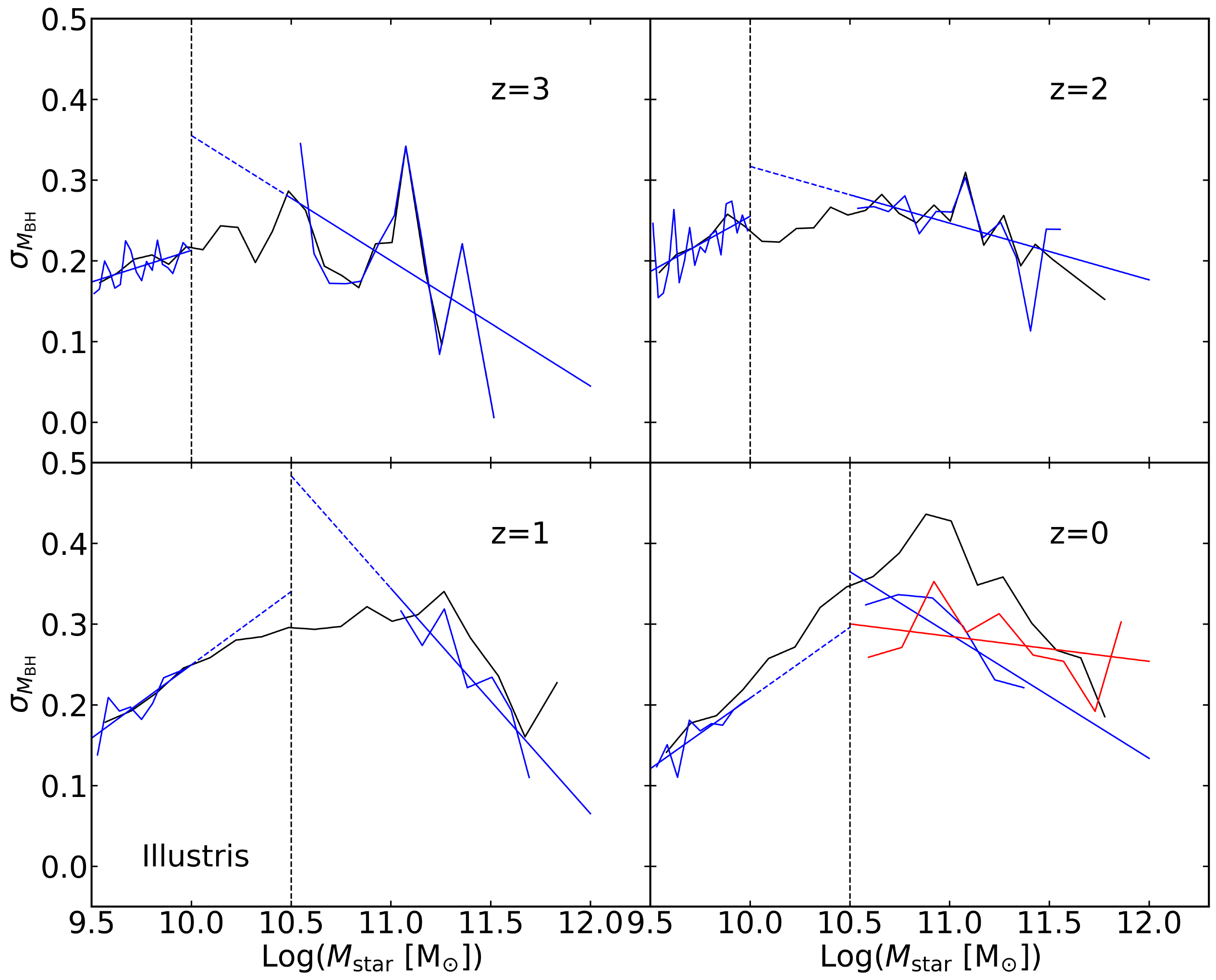}}
   \subfigure{\includegraphics[width=0.45\textwidth]{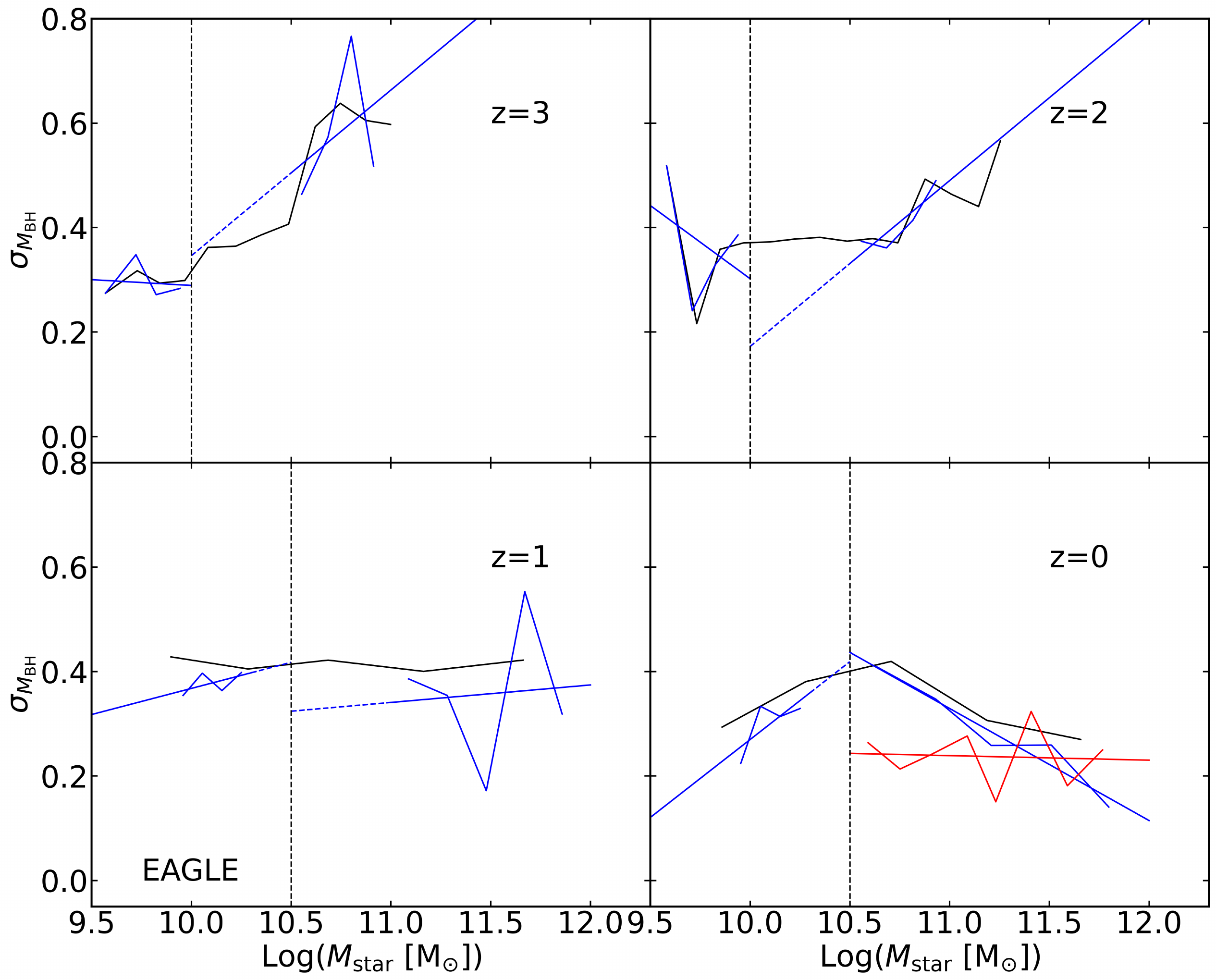}}
   \subfigure{\includegraphics[width=0.45\textwidth]{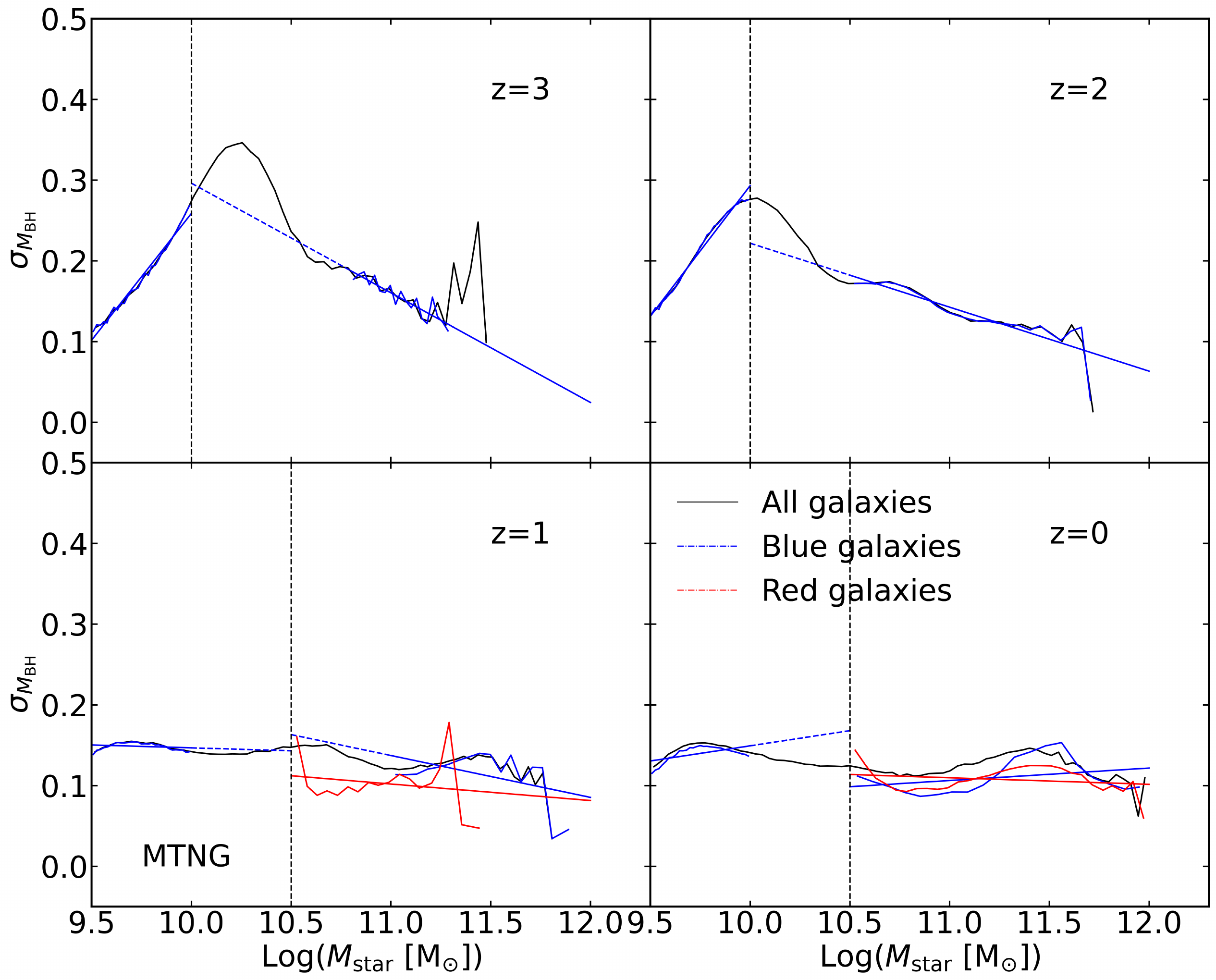}}
   
   \caption{The scatter in the BH mass of galaxies of different stellar mass at different redshifts in the TNG100, Illustris, EAGLE, and MTNG simulations. The black line represents the scatters of BH mass for all galaxies. The blue line gives the scatter only for blue galaxies, while the red line correspondingly is for red galaxies. For the TNG100, Illustris, and MTNG simulations, the blue galaxies are defines as galaxies with color index $(g-r)<0.6$ while the red galaxies are the galaxies with color $(g-r)>0.7$. For EAGLE simulation, the blue galaxies are defined as galaxies with color $(g-r)<0.75$ while the red galaxies are those with color $(g-r)>0.75$. For TNG100, Illustris, and MTNG, there is a color gap since we ignore the green valley galaxies. \textit{Upper-left panel:} TNG100 simulation. \textit{Upper-right panel:} Illustris simulation. \textit{Lower-left panel:} EAGLE simulation. \textit{Lower-right panel:} MTNG simulation. The slightly higher scatter in blue galaxies than in red galaxies at $z=0$ indicates that the scatter contains the information about the growth history of BH. The consistent decrease of scatter with stellar mass at $M_{\star}\gtrsim10^{10.5}\,{\rm M_{\odot}}$ at $z=0$, observed for both red and blue galaxies across all four simulations, suggests that this trend is driven by mergers.}
    \label{sc_mstar_z}
\end{figure*}

\begin{figure*}
   \subfigure{\includegraphics[width=0.49\textwidth]{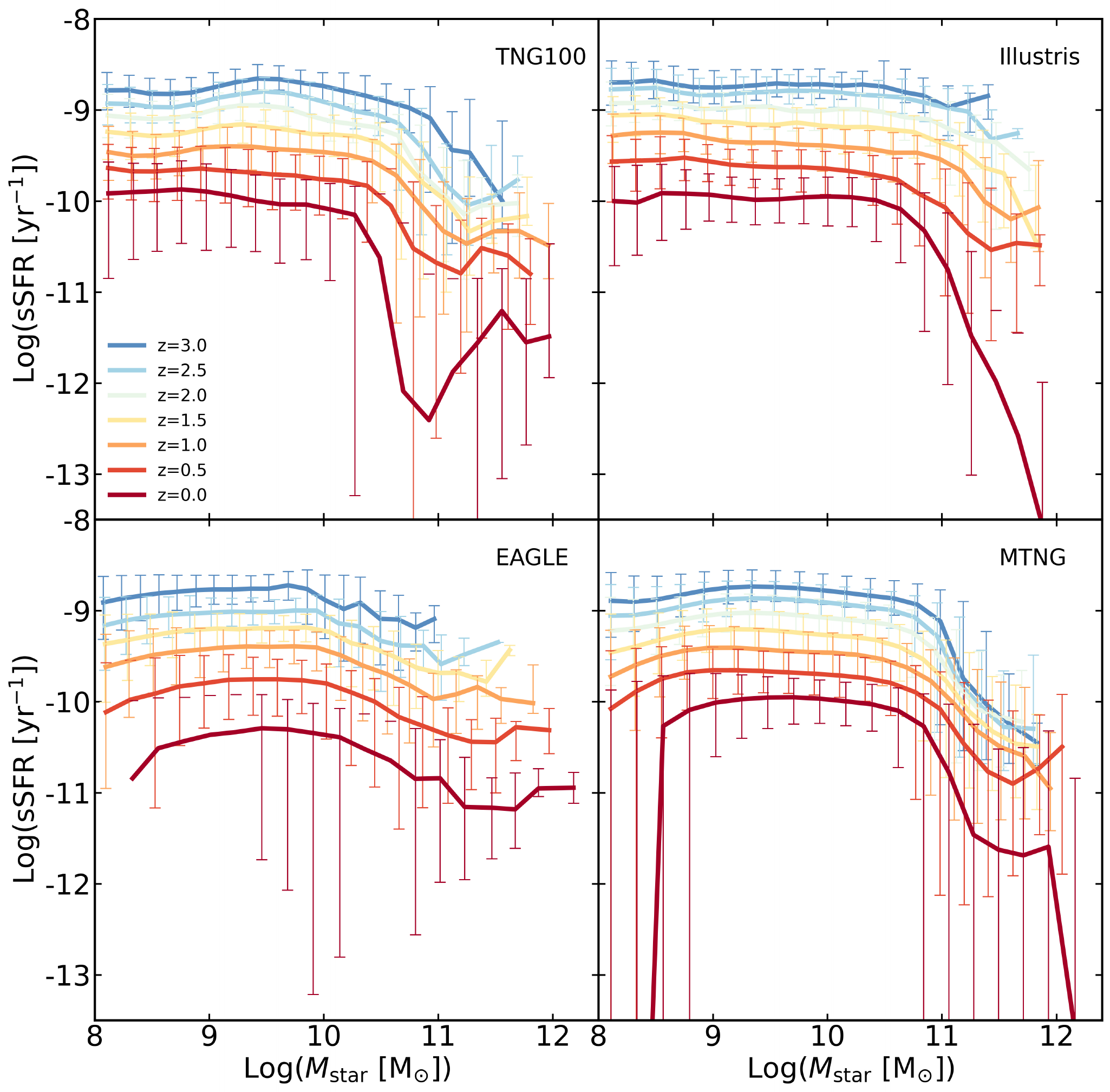}}
   \subfigure{\includegraphics[width=0.49\textwidth]{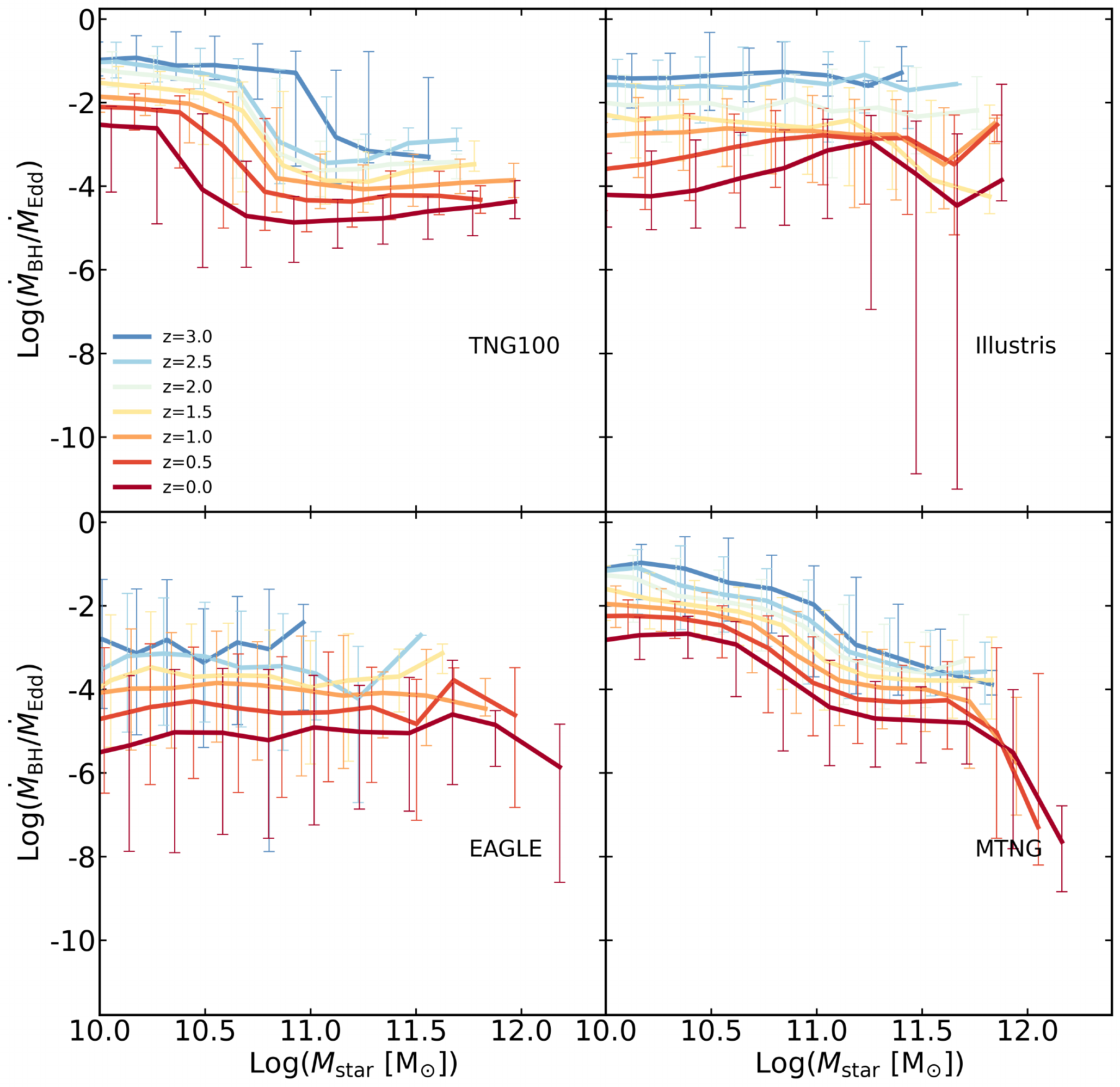}}  
   \caption{\textit{Left panels:} The median value of the specific SFR as a function of stellar mass at different redshifts in the TNG100, Illustris, EAGLE, and MTNG simulations. \textit{Right panels:} The median value of the BHAR as a function of stellar mass at different redshifts in TNG100, Illustris, EAGLE, and MTNG. The significant drop of the sSFR and the BHAR at the high mass end in TNG100/MTNG indicates that the accretion/in-situ star formation become less important for massive galaxies compared to Illustris/EAGLE.}
   \label{sfr_bhar}
\end{figure*}

The result is presented in Figure~\ref{var_results}. The top panel displays the evolution of variance and intrinsic scatter of the BH mass for the massive galaxies in the TNG100 and SeedVar simulations. The intrinsic variance is calculated by assuming that the numerical scatter in SeedVar is similar to that in TNG100. Comparing the resulting intrinsic variance in the SeedVar and TNG100 simulations, the scatter is higher in SeedVar than in TNG100 except for $z\sim0$, where the intrinsic variance in SeedVar is comparable to that in TNG100. The variance in the SeedVar simulation at $z\lesssim 0.5$ experiences a steep decrease, from $\sim0.1$ to $\sim0.01$, i.e.~the intrinsic scatter drops from $\sim0.3$ dex to $\sim 0.1$ dex. This result is likely due to the effective galaxy quenching in TNG simulations.

We further investigate the scatter of BH masses in different BH mass bins in the middle panel of Figure~\ref{var_results}. We found that at different redshifts, the dependence of the scatter of BH masses in different BH mass bins is similar, so we stack the results from $z=3$ to $z=0$. The BHs in SeedVar show high scatter in low-mass BH bins due to the large variation of BH seeds. However, this increase in the scatter quickly decays away at $M_{\rm BH}\sim 10^{8}\,{\rm M_{\odot}}$, which can be interpreted as a loss of memory of the BH seed variations, and becomes $\lesssim 0.1\,{\rm dex}$ for massive BHs. This critical BH mass is the mass at which the impact of AGN feedback becomes significant, and the galaxies begin to be quenched \citep{zinger20}. At this stage, BHs also enter a rapid growth phase, during which their accretion is strongly modulated by AGN feedback. This self-regulated growth helps to suppress the divergence caused by initial seed differences and contributes to the flattening of the scatter distribution \citep[see e.g.][]{McAlpine2018}. Since the quenched galaxies will have relatively low accretion rates due to a lack of gas supply, and the growth of BHs in quenched galaxies is mainly occurring due to mergers, we can infer that the continued drop of the scatter of the BH mass at high masses is primarily due to galaxy quenching and merger-driven averaging at $M_{\rm BH} \sim 10^8\,{\rm M_{\odot}}$ and above.

This transition mass marks the point where AGN feedback becomes significant and galaxies begin to quench \citep{zinger20}. At the same time, BHs enter a rapid growth phase in which accretion is strongly modulated by feedback, leading to self-regulated evolution. This process helps suppress the divergence caused by initial seed differences and contributes to the flattening of the scatter distribution \citep[see e.g.][]{McAlpine2018}.

We also show the time evolution of the fraction of the variance from different origins in the bottom panel of Figure~\ref{var_results}. The figure shows that the BH seed variation is dominating the total scatter from $z=3$ to $z=0$, except at $z\sim 0$, where the variance becomes dominated by the numerical variation and hierarchical merging. The contribution to the variance from hierarchical merging monotonically increases with decreasing redshift, whereas the variance due to the BH accretion is roughly constant at $z>1$, and quickly becomes negligible at $z<1$. The fraction of the variance due to the numerical effects roughly stays at  a constant level from $z=3$ to $z=0$. These results support our interpretation that at high redshift the scatter of the BH masses in the SeedVar simulation is still dominated by the BH seed variation. However, as more and more massive galaxies become quenched, the subsequent growth of BHs becomes dominated by mergers. As a result, the variance due to the variation of BH seeds gradually decreases due to the cessation of BH accretion and the growing importance of dry mergers. At $z=0$, the variance becomes dominated by hierarchical merging again despite the seed mass variations.

\begin{figure}
   \includegraphics[width=0.45\textwidth]{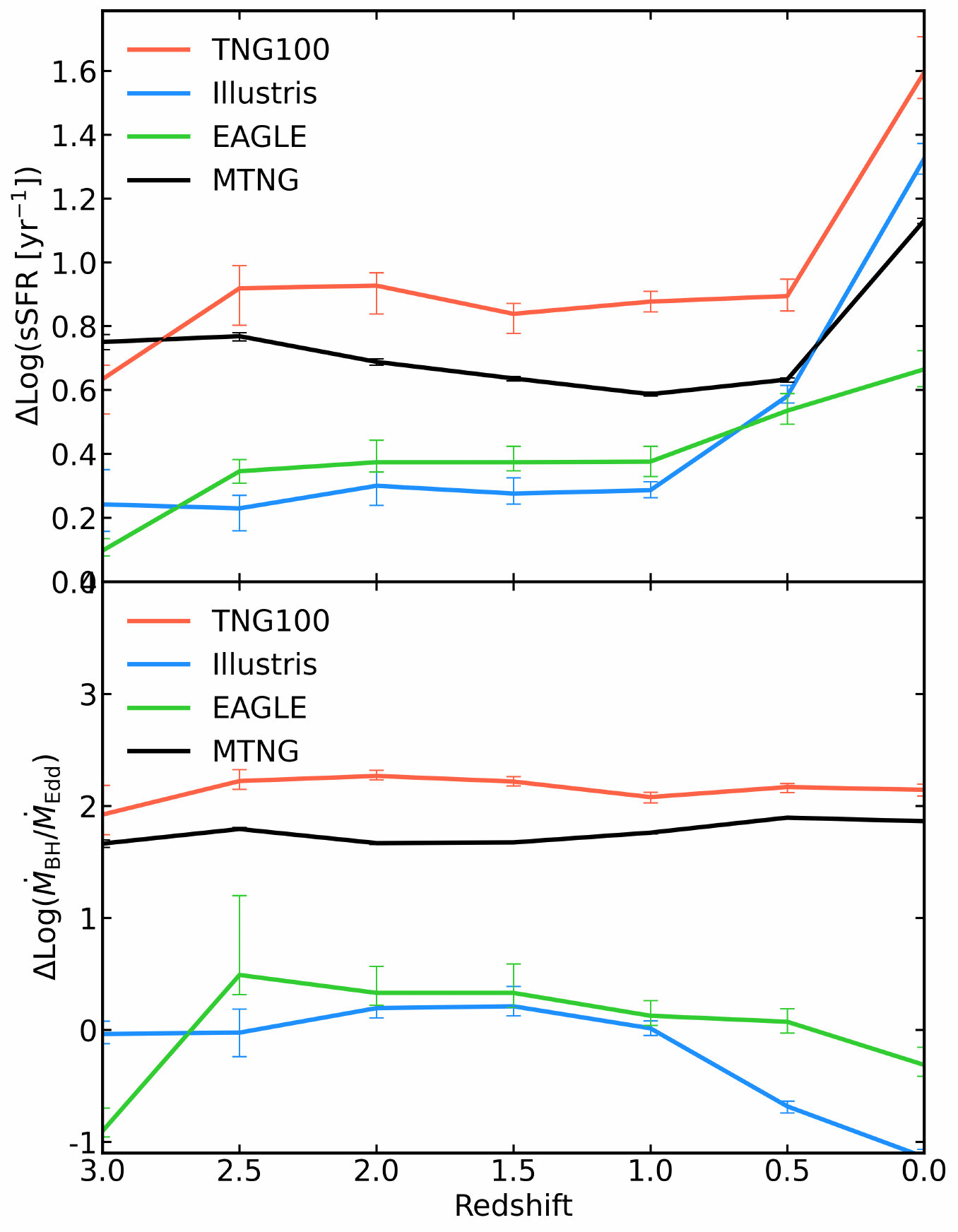}
      \caption{\textit{Upper panel:} The difference of the sSFR for the stellar mass bins in the ranges $10^{10-11}\,{\rm M_{\odot}}$ and $10^{10.5-11.5}\,{\rm M_{\odot}}$ in TNG100, Illustris, EAGLE, and MTNG. \textit{Lower panel:} The difference of the BHAR in units {of} the Eddington rate for the stellar mass bins in the ranges $10^{10-11}\,{\rm M_{\odot}}$ and $10^{10.5-11.5}\,{\rm M_{\odot}}$ in TNG100, Illustris, EAGLE, and MTNG. Similar to Figure \ref{sfr_bhar}, the high difference of sSFR and BHAR in these figures also indicates that the growth of BH mass and stellar mass of massive galaxies proceeds  in TNG100/MTNG mainly through dry mergers instead of in-situ star formation or BH accretion.}
    \label{dsfr_dbhar}
\end{figure}

\section{Discussion}\label{sec:dicussion}

The results of the analysis in the previous section suggest that the scatter in the $M_{\rm BH}-M_{\star}$ relation is related to galaxy evolution. In this section, we discuss the potential of this connection to constrain different models for BH feedback physics.

\subsection{The connection between the scatter and galaxy quenching}

Since the physical origins of the scatter in the BH mass include processes related to hierarchical merging and BH accretion, we also expect it to contain information about galaxy quenching. In Figure~\ref{sc_mstar_z} we probe for this by considering the scatter of the BH mass separately for red and blue galaxies at $z=0$. We find that {in EAGLE} the scatter of the BH mass for red galaxies is significantly lower than for blue galaxies with stellar mass ranging from $10^{10.5}\,{\rm M_{\odot}}$ to $10^{11}\, {\rm M_{\odot}}$, confirming our inference, whereas this difference is at most marginal for the other three simulations.

To further investigate this issue, we display in Figure~\ref{sfr_bhar} the specific star formation rate (sSFR) and  the black hole accretion rate (BHAR) as a function of  stellar mass at different redshifts in the TNG100, Illustris, EAGLE, and MTNG simulations. It can be seen that the sSFR and BHAR experience a significant decrease for massive galaxies in TNG100 and MTNG from $z=3$ to $z=0$. Especially for the BHAR, both TNG100 and MTNG show a sharp drop and the critical stellar mass for the drop decreases with decreasing redshift. For the Illustris and EAGLE simulations, the sSFR is roughly constant for low-mass galaxies and shows a slight decrease at the same redshift compared to TNG100 and MTNG. For the BHAR, the value is roughly constant over the analyzed mass range  at all redshifts in EAGLE, and at $z\gtrsim2$ in Illustris. At $z\lesssim 2$, the BHAR tends to be low for massive galaxies in Illustris. 

We further calculate the differences of the sSFR and BHAR for stellar masses in the ranges $10^{10-11}\,{\rm M_{\odot}}$ and $10^{10.5-11.5}\,{\rm M_{\odot}}$, and display the results in Figure~\ref{dsfr_dbhar}. The upper panel of the figure shows the difference for the sSFR and the lower panel shows the difference for the BHAR in units if the Eddington rate for the two different stellar mass bins in TNG100, Illustris, EAGLE, and MTNG. It can be clearly seen that the differences in TNG100 and MTNG are significantly higher than in Illustris and EAGLE, which is also seen in Figure~\ref{sfr_bhar}, and the trend is more significant for the BHAR. For the sSFR, the difference becomes comparable in the Illustris simulation to the TNG100 and MTNG simulations. The higher difference indicates that the growth of BH mass and stellar mass of massive galaxies proceeds mainly through dry mergers instead of in-situ star formation or BH accretion.

Based on the results presented in Figure~\ref{dsfr_dbhar}, we can draw the following conclusion. Since mergers will significantly reduce the scatter in the $M_{\rm BH}-M_{\star}$ relation and the results in Figure~\ref{Sc_intrinsic} show that the intrinsic scatter of the $M_{\rm BH}-M_{\star}$ relation is lower in TNG100 and MTNG than in Illustris and EAGLE, the results indicate that the galaxy formation models with efficient quenching for massive galaxies will produce a low intrinsic scatter of the $M_{\rm BH}-M_{\star}$ relation for massive galaxies. The BHAR is positively correlated to the sSFR in galaxies since star formation and BH accretion depend on the same gas reservoir in the galaxies. Thus, the scatter in BH mass is correlated to differences in the sSFR and contains information about  galaxy quenching.

\begin{figure}
   \includegraphics[width=0.45\textwidth]{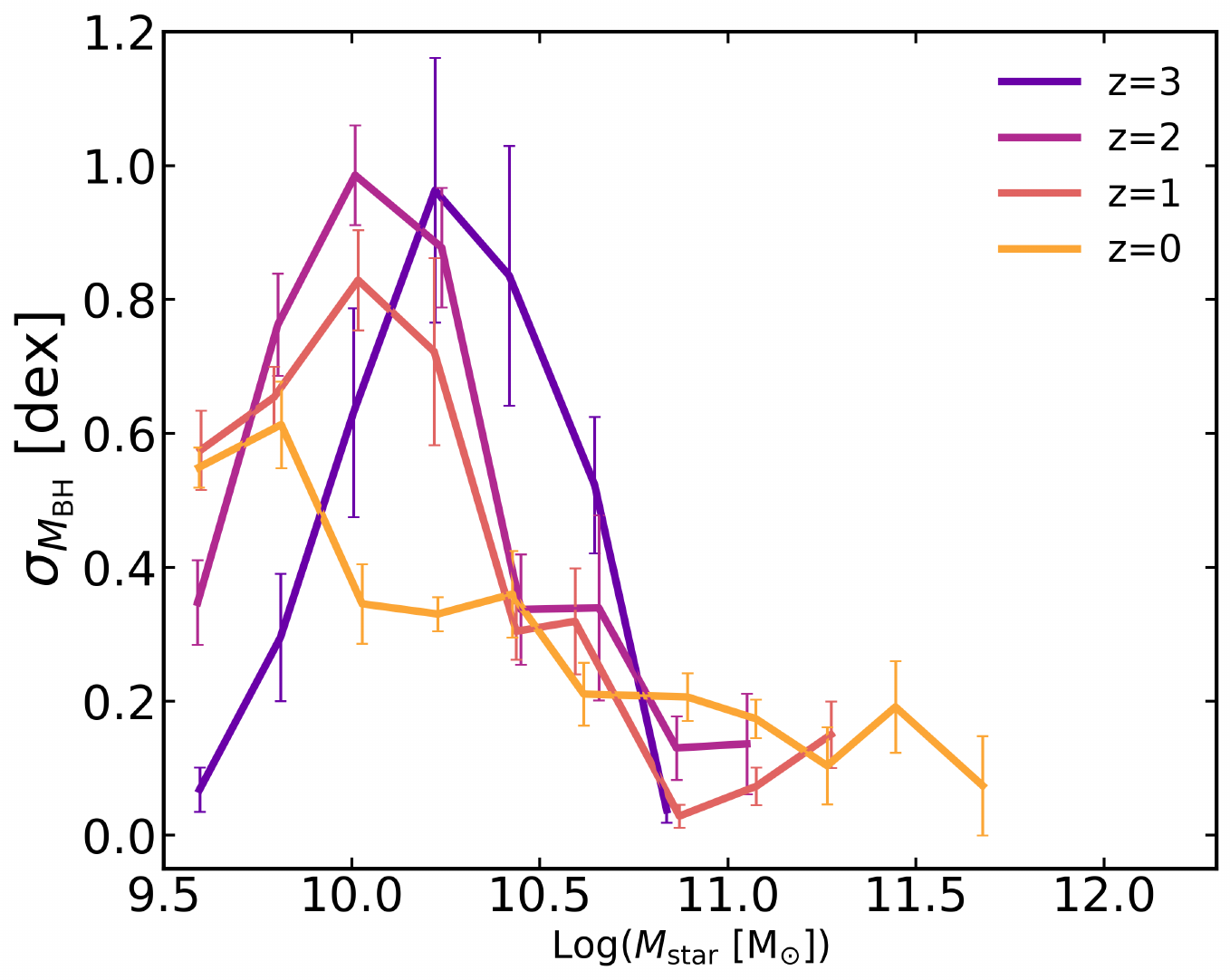}
   
   \caption{The scatter of the BH masses in galaxies as a function of their stellar mass at different redshifts in {our} test simulation with no AGN feedback. Comparing with the results for our four cosmological simulation, the figure suggest that BH feedback regulates the $M_{\rm BH}-M_{\star}$ relation at $M_{\star}\lesssim10^{10.5}\, {\rm M_{\odot}}$, since the scatter for this low mass regime is lower  in the four cosmological simulations than in the NoAGN run.}
    \label{sc_nofb}
\end{figure}

\subsection{Regression to the mean or co-evolution?} \label{sec:RegOrCoev}

Both the hierarchical merging and the self-regulation of BH and stellar mass growth can produce a tight relation between $M_{\rm BH}$ and $M_{\star}$. The former mechanism can be viewed as a consequence of the central-limit theorem. A series of mergers will reduce the scatter in the $M_{\rm BH}-M_{\star}$ relation and regress to a normal distribution around the mean, even if {the} initial distributions of $M_{\rm BH}$ and $M_{\star}$ are uncorrelated.  In contrast, the latter mechanism of self-regulation entails a mutual influence of the growth of BHs and galaxies on each other, through the ability of AGN feedback to regulate star formation, and the tendency of gas cooling to supply fuel to both, star formation and BH accretion. Which of these two mechanisms is mainly responsible for the tight relation of $M_{\rm BH}-M_{\star}$ relation remains under debate.  Since larger scatter indicates weaker correlation, we try to discuss this issue from the viewpoint of the scatter of the $M_{\rm BH}-M_{\star}$ relation.

Figure~\ref{sc_mstar_z} displays the scatter of BH mass in galaxies at different redshifts in the TNG100, Illustris, EAGLE, and MTNG simulations. At first glance we can find, for all simulations at all redshifts, that the scatter for low-mass galaxies exhibits a similar trend: the magnitude always tends to become larger with increasing stellar mass. However, eventually the trend reverses and for high-mass galaxies the scatter becomes again smaller with increasing stellar mass. These opposite trends for low-mass and high-mass galaxies must have different origins, so we analyze them separately.

For the low-mass galaxies, the merger rate is low compared to massive galaxies \citep{rodriguez2015}, and thus the growth of BHs and galaxies is mainly due to in-situ star formation and BH accretion. This implies that the relation between $M_{\rm BH}$ and $M_{\star}$ is mainly regulated by feedback processes. Figure~\ref{sc_mstar_z} shows that the scatter of the BH mass in the Illustris/EAGLE simulations at all redshifts, and for TNG100/MTNG at high redshift, increases with increasing stellar mass in this regime.  For comparison, we also run a small box cosmological simulation with the TNG model but without AGN feedback, see Figure~\ref{sc_nofb}. Since the scatter does not strongly depend on resolution after removing outlier data outside $2\sigma$ (as shown in Section~\ref{sec:res}), we can safely use the results to compare with the other simulations here. From the figure, we find that the scatter can reach $\sim1$ dex when feedback is omitted, which is much higher than the scatter of the BH mass at low-mass galaxies we find for the full simulation. This result indicates that the feedback processes in all four simulation models are important for regulating the star formation and BH accretion even in low-mass galaxies. Also, in the TNG100 and MTNG simulations, the effects of feedback processes are clearly more pronounced than in Illustris/EAGLE, as the scatter in the BH masses can be well controlled at a low level of $\sim0.1-0.2$ dex, as shown in Figure~\ref{sigma&k}.

For massive galaxies, the scatter in the BH masses decreases with increasing stellar mass in almost all simulations at all redshifts, except for MTNG at very low redshift. But in the latter, the scatter of the BH mass in MTNG roughly stays at a constant value of $\sim 0.1-0.15$ dex, which is comparable to the scatter from hierarchical merging presented in Section~\ref{sec:diff_sc}, suggesting that in this case the value already reached the lower limit expected for the TNG model. Similar trends are also found in the simulation without AGN feedback in Figure~\ref{sc_nofb}. These results indicate that the decline of the scatter in the BH mass for massive galaxies is unrelated to the specifics of AGN feedback but rather occurs due to hierarchical merging as a main driver, regardless of the details of the galaxy formation model.

Taken together, our findings indicate that the tight $M_{\rm BH}$–$M_\star$ relation arises from a combination of co-evolutionary processes and statistical convergence via mergers, with the dominant mechanism depending on galaxy mass.

\subsection{Scatter of overmassive BHs} \label{sec:sc_at_noon}

Recent James Webb Space Telescope (JWST) observations find a certain number of overmassive BHs at high redshift \citep{p23, guia2024, mezcua2024}. The mass of these BHs is even comparable to their host galaxies. How they grow remains under debate, and presently represents a significant open question in early galaxy evolution.

We briefly discuss this issue here from the viewpoint of the scatter of the BH masses, focusing on the observations of overmassive BH at cosmic noon \citep{mezcua2024}. Based on \citet{guia2024}, the intrinsic observational scatter is around 0.5 dex for galaxies with stellar mass in the range from $\sim10^{9}\,{\rm M_{\odot}}$ to $\sim10^{9.5}\,{\rm M_{\odot}}$ at $z\sim3$. Our results in this work suggest that such a high scatter may arise from a high scatter of the initial BH seed masses, or be due to a lack of self-regulation in the BH accretion. A further possibility could lie in large observational errors in the BH mass estimates. \citet{bhowmick24} use a physically-motivated BH seeding scheme combined with the TNG model and find that these overmassive BH could grow through mergers. However, since mergers will reduce the scatter, this solution may have problems in reproducing the high scatter that is observed, unless the variation of the initial BH seed mass is extremely high.
This serves to illustrate that  scatter can provide crucial  extra information on the origin of these overmassive BH. In particular, our results appear to indicate that obtaining a high scatter for the overmassive BHs requires a substantial influence of accretion variations and not just mergers alone.

\section{Summary and conclusion}\label{sec:conclusion}

In this work, we studied the scatter in the $M_{\rm BH}-M_{\star}$ relation for massive galaxies predicted by the IllustrisTNG, Illustris, EAGLE, and MillenniumTNG simulations. Our aims have been to characterize the size of this scatter, how it varies with stellar mass and redshift, and to elucidate numerical and physical contributions to it, as well as possible connections to galaxy evolution and AGN feedback processes.

To this end we have complemented the simulation data with several small-box cosmological test simulations, in particular to isolate the influence of numerical noise and of BH merging on the scatter.  We have addressed this by using five cosmological simulations with the full TNG model and identical initial conditions but different parallelization settings (which introduces differences in round-off and random number usage), and five cosmological simulations where BH accretion was turned off. With the help of these simulations, we could distinguish quantitatively between different origins of the scatter. Finally, we have discussed the possible connections between the scatter in the $M_{\rm BH}-M_{\star}$ relation and three interesting aspects of galaxy evolution, namely the galaxy quenching phenomenon, the co-evolution of galaxies and BHs, and the apparent existence of overmassive BHs that are far away from the $M_{\rm BH}-M_{\star}$ relation. Our main findings can be summarized as follows:

\begin{itemize}

\item {As shown in Figure \ref{sigma&k}}, with respect to the scatter and the slope of the $M_{\rm BH}-M_{\star}$ relation for massive galaxies, the TNG100, Illustris, EAGLE, and MTNG simulations can be divided into two groups: TNG100/MTNG and Illustris/EAGLE. TNG100 and MTNG show a similar evolution of the scatter and the slope, with the scatter being roughly $0.1-0.2$ dex and the slope ranging from 1 to 0.75 at $z=3\sim0$. As expected, MTNG is very close {to} TNG. In contrast, Illustris and EAGLE exhibit a higher scatter across $z=3$ to $z=0$, whose value is $\sim0.3-0.5$ dex, and the slope of the $M_{\rm BH}-M_{\star}$ relation tends to be steeper than 1. This result is insensitive to the selection criteria of galaxies, and is unlikely to be particularly sensitive to resolution (as shown in Figure \ref{sigma&k} and \ref{res_sigma&k}), rather it reflects more fundamental differences in the way BH feedback is treated in the simulations. We also note that previous work \citep[e.g.,][]{bahe2022} has shown that changes in repositioning implementations can enhance SMBH merger rates and thus in principle alter the scatter of the $M_{\star}$-$M_{\rm BH}$ relation.

\item The sources of the scatter can be broadly divided into numerical and physical origins. The numerical origins can be further divided into being due either to numerical round-off noise or the use of stochastic modelling of  subgrid physics. With respect to physical origins, we here identified hierarchical merging, BH accretion, and possible variations of the seed BH mass as main contributions. Through calculating the variance of matched halos in two sets of five cosmological simulations with identical initial conditions where in one the full TNG model including BH accretion was used, and in another BH accretion was turned off, we could separately estimate the scatter from numerical sources. We find that the cleaned variance (with $3\sigma$ outliers removed) from numerical variations (introduced by floating point round-off differences and stochastic prescriptions) is around $0.01$ (the red line in Figure \ref{BHnumscev}), which constitutes more than $\sim50\%$ of the total variance at $z<1.5$ (the red line in the bottom panel of Figure \ref{TNGscComponent}).

\item We also estimated the scatter from  hierarchical merging in the simulations with the TNG model by turning off BH accretion. The corresponding variance increases with time and reaches $\sim 0.01$ at $z=0$ (the black line in the upper panel of Figure \ref{TNGscComponent}). 

\item By subtracting the scatter from numerical origins, we have also obtained estimates for the intrinsic physical scatter predicted by the simulation models. {Note that the comparison to observational constraints is performed at $z = 0$ only.} We find that the intrinsic scatter in Illustris and EAGLE is nominally close to the scatter seen in observations (See the upper panel of Figure \ref{Sc_intrinsic}). Note, however, that the latter can potentially have substantial contributions from measurement errors, as discussed in \citet{kormendy2013}, the measurement errors are estimated to be $\sim0.1-0.2$ dex, which can constitute a substantial fraction of the total observed scatter ($\sim$0.34 dex). For Illustris and EAGLE to be viable, these errors would have to be quite small. Only if the scatter had been significantly larger in the simulations than in the observations, there would be a clear tension. Assuming the scatter from hierarchical merging is similar in all four simulations, the scatter due to BH accretion can also be estimated. We find that the scatter from BH accretion dominates the scatter in the Illustris and EAGLE simulations while it becomes negligible at $z\lesssim 1$ in the TNG100 and MTNG simulations (the lower panel of Figure \ref{Sc_intrinsic}), suggesting that the growth of black holes in massive galaxies is mainly dominated by dry merging in TNG100 and MTNG at late times.

\item Variations of the BH seed mass can also contribute to the scatter in the $M_{\rm BH}-M_{\star}$ relation. While all fiducial simulations adopt uniform seed masses, we have demonstrated this explicitly with a test simulation where we imposed a 0.5 dex scatter in the BH seed masses (see Figure~\ref{var_results}). In this case, the scatter of the $M_{\rm BH}-M_{\star}$ relation can reach up to $\sim 1$ dex in a TNG-like simulation. However, the scatter then still decreases substantially at $z<0.5$, where it becomes similar to the default TNG simulation. This can be understood as a result of the effective quenching of massive galaxies in the TNG model, so that hierarchical merging becomes effective in reducing the scatter in more massive galaxies at late times. Feedback models like Illustris and EAGLE would thus be expected to keep a memory of the initial scatter in the BH seed masses for a longer time.

\item Our results have demonstrated that the scatter of the $M_{\rm BH}-M_{\star}$ relation for massive galaxies is closely related to the ability of different BH models to regulate the gas availability and thereby quench massive galaxies. We argue that Illustris and EAGLE show a higher scatter because they quench massive galaxies somewhat less resolutely than {TNG100} and MTNG, due to less effective removal or heating of gas. This is also the reason why we find that the scatter from BH accretion is the dominating contribution  in the Illustris and EAGLE models.

\end{itemize}

In sum, our results have demonstrated that the amplitude and slope of the $M_{\rm BH}-M_{\star}$ relation are not the only relevant constraints for galaxy formation models that account for supermassive BHs. A similarly interesting quantity is the scatter around this relation, which is closely related to various aspects of the physics of the joint evolution of galaxies and their embedded supermassive BHs. Our findings here have demonstrated that different models for the initial BH seeding mechanism, the subsequent gas accretion, and the impact of BH feedback on BH growth and galaxy quenching all leave subtle imprints in the magnitude of the scatter and its evolution, as well as its trends with stellar mass. This could well turn into a decisive test for distinguishing different models, provided different sources of scatter -- including the measurement uncertainties in observations and the numerical noise in simulations -- become sufficiently well understood.

\section*{Acknowledgements}

We thank the anonymous referee for their very insightful and constructive comments that helped to improve the initial manuscript. The authors acknowledge support by a Max Planck Partner group between the Shanghai Astronomical Observatory and the Max Planck Institute for Astrophysics (MPA). Computations were performed on the Freya compute cluster of MPA, operated by the Max Planck Computing and Data Facility.

%%%%%%%%%%%%%%%%%%%%%%%%%%%%%%%%%%%%%%%%%%%%%%%%%%
\section*{Data Availability}

The data of this study are available from the corresponding author upon reasonable request.



\bibliographystyle{mnras}
\bibliography{ref} % if your bibtex file is called example.bib

% 

% Don't change these lines
\bsp	% typesetting comment
\label{lastpage}
\end{document}